\documentclass[12pt]{article}
\usepackage{xcolor}
\usepackage{bm}
\usepackage{amsmath,amsfonts,amsthm}
\usepackage{graphicx,psfrag,epsf}
\usepackage{enumerate}
\usepackage{natbib}

\usepackage{siunitx}

\newcommand{\proj}{\mbox{proj}}

\usepackage[noend]{algpseudocode}
\usepackage{algorithm}
\usepackage{algorithmicx}
\newcommand*\Let[2]{\State #1 $\gets$ #2}
\algrenewcommand\algorithmicrequire{\textbf{Input:}}
\algrenewcommand\algorithmicensure{\textbf{Input:}}
\usepackage{pifont}
\definecolor{cadmiumgreen}{rgb}{0.0, 0.42, 0.24}
\definecolor{caribbeangreen}{rgb}{0.0, 0.8, 0.6}
\definecolor{chartreuse}{rgb}{0.5, 1.0, 0.0}
\definecolor{blue-green}{rgb}{0.0, 0.87, 0.87}
\definecolor{cyan}{rgb}{0.0, 1.0, 1.0}

\newcommand{\red}[1]{\textcolor{black}{#1}}

\newcommand{\blue}[1]{\textcolor{black}{#1}}
\newcommand{\cmark}{\blue{\ding{51}}}%
\newcommand{\xmark}{\red{\ding{55}}}%
\newcommand{\LimeTr}{\texttt{LimeTr} }%

\newtheorem{theorem}{Theorem}
\newtheorem{remark}{Remark}
\newtheorem{corollary}{Corollary}

\newtheorem{definition}{Definition}

\newtheorem{assumption}{Assumption}

\begin{document}

\def\spacingset#1{\renewcommand{\baselinestretch}%
{#1}\small\normalsize} \spacingset{1}


  \title{\bf Trimmed Constrained Mixed Effects Models: Formulations and Algorithms}
  \author{Peng Zheng$^\dagger$, Ryan Barber$^\dagger$, Reed Sorensen$^\dagger$, \\Christopher Murray$^\dagger$, and Aleksandr Aravkin$^{\circ\dagger}$
  \thanks{
    This work was funded by the Bill \& Melinda Gates Foundation. 
    The authors also gratefully acknowledge the Washington Research Foundation Data Science Professorship. 
    \red{
    The authors are very grateful to the referees for  insightful questions and suggestions that have improved exposition and extended the scope of the paper. }
    }\hspace{.2cm}\\\\
    \small
    Institute for Health Metrics and Evaluation, University of Washington$^\dagger$ \\
    \small
        Department of Applied Mathematics, University of Washington$^\circ$}
  \maketitle

\bigskip
\begin{abstract}
Mixed effects (ME) models inform a vast array of problems in the physical and social sciences, 
and are pervasive in meta-analysis. We consider ME models where the 
random effects component is linear. 
We then develop an efficient approach for a broad problem class that allows nonlinear measurements, priors, and constraints, 
and finds robust estimates in all of these cases using trimming in the associated marginal likelihood. 

The software accompanying this paper is disseminated as an open-source Python package called \LimeTr. 
\blue{ \LimeTr  is able to recover results more accurately in the presence of outliers} compared to available packages for both standard longitudinal analysis and meta-analysis, 
\blue{and is also more computationally efficient than competing robust alternatives.}
\red{Supplementary materials that reproduce the simulations, as well as run \LimeTr and third party code are available online.}
We also present analyses of global health data, where we use advanced functionality of  \LimeTr , including
 constraints to impose monotonicity and concavity for dose-response 
relationships. Nonlinear observation models allow new analyses in place of classic 
approximations, such as log-linear models. Robust extensions in all analyses ensure that spurious data points 
do not drive our understanding of either mean relationships or between-study heterogeneity. 

\end{abstract}

\noindent%

\spacingset{1.45}
\section{Introduction}
\label{sec:intro}
Linear mixed effects (LME) models play a central role in a wide range of analyses~\citep{DouglasBates2015FLMM}.  
Examples include longitudinal analysis~\citep{laird1982random}, meta-analysis~\citep{dersimonian1986meta}, and numerous 
domain-specific applications~\citep{zuur2009mixed}. 

Robust LME models are typically obtained by using heavy tailed error models for random effects. The Student's t distribution~\citep{pinheiro2001efficient}, as well as 
weighting functions~\citep{koller2016robustlmm} have been used. The resulting formulations are fit either by EM methods, 
estimating equations, or by MCMC~\citep{rosa2003robust}. 
In this paper, we take a different track,  and extend the least trimmed squares (LTS) method to the ME setting. 
\blue{While LTS has found \blue{wide} use in  a range of applications~\citep{AravkinDavis,yang2015robust,yang2018general},
trimming the ME likelihood extends prior \red{work}}.

{\bf Contributions.}
\red{In this paper, we consider a subclass of nonlinear mixed effects models.}  
We allow nonlinear measurements, priors, and constraints, but require that the random effects enter the model in a linear way.  
\blue{We call this class {\it partially nonlinear} ME models, and it covers a broad class of problems while allowing tractable 
algorithms.} 
\blue{We develop new conditions that guarantee the existence of estimators for partially nonlinear models, a trimming approach 
that  robustifies any linear or partially nonlinear model against outliers, and algorithms for solving the  nonconvex optimization problems required to 
find estimates with standard guarantees (convergence to stationary points).
 We also show splines (and associated shape constraints) can be used to capture key nonlinear relationships, 
and illustrate the full modeling capability on real-data examples based on dose-response relationships.  
}

\begin{table}[h!]
\caption{\label{table:novelty} Comparison with currently available robust mixed effects packages. }
\begin{tabular}{|c|c|c|c|c|c|c|} \hline
& \LimeTr & \texttt{metafor}& \begin{tabular}{c} \texttt{robumeta}\\ \texttt{metaplus}\end{tabular}& \begin{tabular} {c} \texttt{robustlmm}\\ \texttt{heavy}\end{tabular}  & \texttt{clme}&\texttt{INLA}\\ \hline
Robust option & \cmark & \xmark & \cmark &\cmark & \xmark & \cmark \\ \hline
\begin{tabular}{c}
Allows for known \\ observation variance
\end{tabular}  & \cmark & \cmark & \cmark & \xmark & \xmark & \cmark\\ \hline 
\begin{tabular}{c}
Covariates in random \\  effects variance 
\end{tabular}& \cmark & \xmark & \xmark & \cmark &\cmark  & \xmark \\ \hline  
Nonlinear observations & \cmark & \xmark & \xmark & \xmark &  \xmark & \cmark \\ \hline 
Linear constraints & \cmark & \xmark & \xmark & \xmark &  \cmark & \xmark \\ \hline 
Nonlinear constraints & \cmark & \xmark & \xmark & \xmark &  \xmark & \xmark \\ \hline 
\end{tabular}
\end{table}

\noindent

The main code to perform the inference is published as open source Python package called \LimeTr  (Linear Mixed Effects with Trimming, pronounced {\it lime tree}). All synthetic experiments using \LimeTr  have been submitted for review as supplementary material with this paper. 
The \LimeTr  package allows functionality that is not available through other available open source tools. The functionality of \LimeTr 
is summarized in Table~\ref{table:novelty}.

The paper proceeds as follows. In Section~\ref{sec:basics}, we describe the problem class of ME models and 
derive the marginal maximum likelihood (ML) estimator. 
In Section~\ref{sec:con}, we describe how constraints and priors are imposed on parameters. 
In Section~\ref{sec:trim}, we review trimming approaches and develop a new trimming extension for the ML approach.
In Section~\ref{sec:algo}, we present a customized algorithm based on variable projection, along with a convergence analysis. 
In Section~\ref{sec:spline}, we discuss spline models for dose-response relationships and give examples of 
shape-constrained trimmed spline models. 
Section~\ref{sec:verify} shows the efficacy \blue{of} the methods for synthetic and empirical data. 
In Section~\ref{sec:synth}, we validate
the ability of the method to detect outliers when working with heterogeneous longitudinal data, and compare with other packages. 
In Section~\ref{sec:real} we apply the method to analyze empirical data sets for both linear and nonlinear relationships using trimmed constrained MEs. This section highlights new capability of \LimeTr  that is not available in other packages.

\section{Methods}
\label{sec:meth}

\subsection{\blue{Notation and Modeling Concepts}}

\blue{In this section, we define notation and concepts used throughout the paper. Additional definitions and notation are introduced in the analysis section. 
We use lower case letters to denote scalars, e.g. $\beta$, and scalar-valued functions, e.g. $f(\beta)$, and bold letters represent vectors, e.g. $\bm \beta$, and vector-valued functions, e.g. $\bm f_i(\bm\beta)$. We use capital bold letters to represent matrices, e.g. $\bm X$. All variables and vectors are real, i.e. in $\mathbb{R}^k$, with $k$ indicating dimension.
For a smooth function $f: \mathbb{R}^{k} \rightarrow \mathbb{R}$, we denote the vector of first derivatives, or {\bf gradient}, by $\nabla f$, and the 
matrix of second derivatives, or {\bf Hessian}, by $\nabla^2 f$.
We use $\mbox{diag}(\bm x)$  to denote a diagonal matrix whose diagonal is an input vector $\bm x$. 
 We use $\bm M^{-1}$ to denote the inverse of a matrix, and denote weighted norms (Mahalanobis distances) by 
\[
\|\bm x\|_{\bm M}^2 := \bm x^T \bm M \bm x,
\]
and \red{$|\bf M|$ to denote the determinant of $\bf M$}.  We use the $\odot$ notation to denote the Hadamard product or operation, so in particular
\[
\bm x \odot \bm y  = \begin{bmatrix} x_1 y_1 \\ \vdots \\ x_n y_n \end{bmatrix}, \quad \bm x^{\odot \bm y}  = \begin{bmatrix} x_1^{y_1} \\ \vdots \\ x_n^{y_n} \end{bmatrix}.
\]
A {\bf likelihood}  \red{maps parameters to an associated density function for observed data. In the mixed effects context, these parameters can be separated into {\bf fixed effects} (e.g. population mean) and {\bf random effects} (e.g. study-specific random intercept).} 
A {\bf marginal} likelihood function refers to the likelihood obtained by integrating out random effects from the joint likelihood.  
} 

\blue{
 \red{We  incorporate additional information about parameters using statistical priors, and restrict parameter domains using {\bf constraints}}. When constraints are present we use the term {\bf constrained likelihood}.  
We use {\bf trimming} ( Section~\ref{sec:trim}) to robustify a (marginal) likelihood, and use the term {\bf trimmed likelihood} to describe such likelihoods.}

\blue{
The goal of an inference problem is to maximize the (marginal) likelihood or modified likelihood, or equivalently to minimize the negative logarithm of such a function, \red{called an {\bf objective function} in optimization.} The optimization problem specification includes constraints as well as the objective. 
We use $\min$ to refer to the minimum value of an optimization problem, and $\arg\min$ to refer to the minimizer, which 
corresponds to the {\bf estimator} \red{in this setting}.  
}

\blue{
We define projection of a point $\bm x $ onto a closed set $C$ by 
\[
\red{\proj}_C(\bm x) := \arg\min_{\bm y \in C} \frac{1}{2}\|\bm y-\bm x\|^2.
\]
 }

\subsection{Problem Class }
\label{sec:basics}

We consider the following mixed effects model:
\begin{equation}
\label{eq:coords}
\begin{aligned}
\bm y_{i} & = \bm f_i(\bm \beta)  + \bm Z_i \bm u_i +\bm  \epsilon_{i}, \quad \blue{i = 1, \dots, m}, \\
\bm u_i & \sim N(\bm 0, \bm \Gamma ), \quad \bm \Gamma = \mbox{diag}(\bm \gamma), \quad \bm \epsilon_{i}  \sim N(\bm 0,\bm \Lambda_i), 
\end{aligned}
\end{equation}
\blue{where $m$ is the number of groups,} $\bm y_i \in \mathbb{R}^{n_i}$ is the vector of observations from the $i$th group, 
\blue{ and $n = \sum_i n_i$ is the total number of observations.} 
\blue{Measurement errors are denoted by $\bm\epsilon_i \in \mathbb{R}^{n_i}$}, with covariance $\bm \Lambda_i \blue{\in\mathbb{R}^{n_i \times n_i}}$, 
\blue{and we denote by $\bm \Lambda\in \mathbb{R}^{n\times n}$ the full block diagonal measurement error covariance matrix.}
\blue{Regression coefficients are denoted by} $\bm \beta \blue{\in \mathbb{R}^{k_\beta}}$.
\blue{Random effects are denoted by} $\bm u_i \in \mathbb{R}^{k_{\bm \gamma}}$, 
where $\bm Z_i \in \mathbb{R}^{n_i \times k_{\bm \gamma}}$ are linear maps. 
The \blue{functions} $\bm f_i \blue{: \mathbb{R}^{k_\beta} \rightarrow \mathbb{R}^{n_i}}$ may be nonlinear, but we restrict the random effects 
to enter in a linear way through the \blue{linear maps $\bm Z_i: \mathbb{R}^{k_\gamma}\rightarrow \mathbb{R}^{n_i}$}. 

 A range of assumptions may be placed on $\bm \Lambda$. 
In longitudinal analysis, $\bm \Lambda$ is often a diagonal or block-diagonal matrix, 
parametrized by \blue{a small set of parameters, with the simplest example $\bm \Lambda = \sigma^2 \bm I$ where $\sigma^2$ is unknown}. 
In meta-analysis, $\bm \Lambda$ is a known diagonal matrix 
whose entries are variances for each input datum. 
\blue{We do not restrict the term `meta-analysis' to a single observation per study, since 
many analyses include multiple observations, such as summary results for quartiles based on exposure. The distinguishing feature 
of meta-analysis is the specification of a known $\bm \Lambda$ matrix.}

\blue{For convenience, we denote by $\bm \theta$ the tuple of fixed parameters: 
\[
\bm \theta :=(\bm \beta, \bm \gamma, \bm \Lambda).
\]
}
The joint likelihood \blue{corresponding to model~\eqref{eq:coords}} for $\bm \theta$ and random effects $\bm u$ 
is given by 
 \begin{equation}
 \label{eq:ML} 
p(\bm \theta, \bm u | \bm y) =
\frac{\exp\left(-\frac{1}{2}\|\bm u\|^2_{\bm \Gamma^{-1}}\right)}{\sqrt{|2\pi \bm \Gamma|}}\prod_{i=1}^m
\frac{\exp\left(-\frac{1}{2}\|\bm y_{i} - \bm f_i(\bm \beta) - \bm Z_i \bm u\|^2_{\bm \Lambda^{-1}}\right)}{\sqrt{|2\pi \bm \Lambda_i|})}
 \end{equation}
\blue{Maximizing~\eqref{eq:ML} with respect to both fixed and random parameters is problematic, as the number of random parameters $\bm u_i$ grows 
with the number of groups. In the extreme case of one observation per group, there are more unknowns than datapoints. 
Standard practice is to marginalize random effects, integrating~\eqref{eq:ML} with respect to all $\bm u_i$. The numerical estimation is then accomplished 
by taking the negative logarithm (a simplifying transformation) and minimizing the result, which is an equivalent problem to maximizing the marginal likelihood:}
  \begin{equation}
 \label{eq:MML}
 \begin{aligned}
 &\mathcal{L}_{ML} \blue{(\bm \theta)}  =  -\ln\left(\int p(\blue{\bm \theta}, \bm u | \bm y) d\bm u\right) \\
& \propto \sum_{i=1}^m  
 \frac{1}{2}(\bm y_i - \bm f_i (\bm \beta))^\top (\bm Z_i \bm \Gamma\bm Z_i^\top + \bm \Lambda_i )^{-1}(\bm y_i - \bm f_i(\bm \beta)) 
 + \frac{1}{2}\ln|\bm Z_i\bm \Gamma\bm Z_i^\top + \bm \Lambda_i  |.
\end{aligned}
 \end{equation}
Problem~\eqref{eq:MML} is equivalent to a maximum likelihood formulation arising from a Gaussian model with correlated errors:  
 \[
 \bm y_i = \bm f_i (\bm \beta) + \bm \omega, \quad \bm \omega \sim N(\bm 0, \bm Z_i \bm \Gamma\bm Z_i^\top + \bm \Lambda_i ).
 \]
The structure of this objective depends on the structural assumptions on $\bm \Lambda$.
In the scope of this paper we always assume $\bm \Lambda$ is a diagonal
matrix, namely all the measurement error are independent with each other.
 We restrict our numerical experiments to two particular classes: (1) $\bm \Lambda = \sigma^2 \bm I$ with 
$\sigma^2$ unknown, used in standard longitudinal analysis, and  (2) the measurements variance is provided and 
used in meta-analysis. 
 \LimeTr  allows other structure options as well, for example 
group-specific unknown $\sigma_i^2$ that extends case (1), but we consider a simple
set of synthetic results to help focus on robust capabilities.

\subsection{Constraints and Priors}
\label{sec:con}
The ML estimate~\eqref{eq:MML} can be extended to incorporate linear and nonlinear inequality constraints
\[
\bm C(\bm \theta)  \leq \bm c,
\]
where $\bm \theta$ are any parameters of interest.  Constraints play a key role in Section~\ref{sec:spline}, 
when we use polynomial splines to model nonlinear relationships. The trimming approach developed in  
the next section is applicable to both constrained and unconstrained ML estimates.  

In many applications it is essential to allow priors on parameters of interest $\bm \theta$. We assume that priors follow a distribution 
defined by the density function   
\[
\bm \theta \sim \exp(-\rho(\bm\theta))
\] 
where $\rho$ is smooth (but may be nonlinear and nonconvex). The likelihood problem is then augmented by adding the term $\rho(\bm \theta)$
to the ML objective~\eqref{eq:MML}. The most common use case is $\rho(\cdot) = \frac{1}{2\sigma_p^2}\|\cdot\|^2$, 
for some user-defined $\sigma_p^2$.

In the next section we describe trimmed estimators, and extend them to the ME setting.

\subsection{Trimming  in Mixed Effect Models}
\label{sec:trim}

Least trimmed squares (LTS) is a robust estimator proposed by~\cite{rousseeuw1985multivariate,rousseeuw1993alternatives} for the standard regression problem. 
\blue{Starting from a standard least squares estimator,}
\begin{equation}
\label{eq:LTS}
\min_{\bm \beta}  \sum_{i=1}^n \frac{1}{2}(y_i - \langle \bm x_i, \bm \beta\rangle)^2,
\end{equation}
the LTS \blue{approach modifies~\eqref{eq:LTS} to minimize} the sum of  {\it smallest $h$} residuals rather than all residuals. 
These estimators were initially introduced to develop linear regression estimators that have a high breakdown point (in this case 50\%) and good statistical efficiency (in this case $n^{-1/2}$).\footnote{Breakdown refers to the percentage of outlying points which can be added to a dataset before the resulting M-estimator can change in an unbounded way. Here, outliers can affect both the outcomes and training data.} LTS estimators are robust  against outliers, and arbitrarily large deviations that are trimmed 
do not affect the final \blue{estimate}.


\blue{The explicit LTS extension to~\eqref{eq:LTS}  is formed by introducing} auxiliary variables $\bm w$:
\begin{equation}
\label{eq:LTSw}
\min_{\bm \beta, \bm w}  \sum_{i=1}^n w_i\left(\frac{1}{2} (y_i - \langle \bm x_i, \bm \beta\rangle)^2\right)
 \quad \mbox{s.t.} \quad \bm 1^\top \bm w = h, \quad \bm 0 \leq \bm w \leq \bm 1.
\end{equation}
The set 
\begin{equation}
\label{eq:cappedSimplex}
\Delta_h := \left\{ \bm w:  \bm 1^\top \bm w = h, \quad \bm 0 \leq \bm w \leq \bm 1 \right\}
\end{equation}
is known as the {\it capped simplex}, since it is the intersection of the $h$-simplex with the unit box~(see e.g.~\cite{AravkinDavis} for details). 
For a fixed $\bm \beta$, the optimal solution of~\eqref{eq:LTSw} with respect to  $\bm w$ assigns weight $1$ to each of the smallest $h$ residuals, and $0$ to the rest. 
Problem~\eqref{eq:LTSw} is solved {\it jointly} in $(\bm \beta, \bm w)$, simultaneously finding the regression estimate and classifying the observations into inliers and outliers. 
This joint strategy makes LTS different from post hoc analysis, where a model is first fit with all data, and then outliers are detected using that estimate. 

Several approaches for finding LTS and other trimmed M-estimators have been developed, including 
FAST-LTS~\citep{rousseeuw2006computing}, \blue{and} exact algorithms with exponential complexity~\citep{mount2014least}.  The LTS approach~\eqref{eq:LTSw} does not depend on the form of the least squares function, and this insight has  been used to extend LTS to a broad range of estimation problems, including generalized linear models~\citep{neykov2003breakdown}, 
high dimensional sparse regression~\citep{alfons2013sparse}, and graphical lasso~\citep{yang2015robust,yang2018general}. The most general problem class to date,  
presented by~\cite{AravkinDavis}, is formulated as 
\begin{equation}
\label{eq:GLTSw}
\min_{\bm \beta, \bm w}  \sum_{i=1}^n w_if_i(\bm \beta) + R(\bm \beta)
 \quad \mbox{s.t.} \quad \bm 1^\top \bm w = h, \quad \bm 0 \leq \bm w \leq \bm 1.
\end{equation}
where $f_i$ are continuously differentiable (possibly nonconvex) functions and $R$ describes any regularizers and constraints (which may also be nonconvex).  

 \red{Problem~\eqref{eq:MML} is not of the form~\eqref{eq:GLTSw}, except for the special case where we want to detect  {\it entire outlying groups}}. 
This is limiting, since we want to differentiate measurements within groups.  
We solve the problem by using a new trimming formulation.

To explain the approach we focus on trimming a single group term from the likelihood~\eqref{eq:MML}:
\[
\left( \frac{1}{2}(\bm y_i - \bm f_i (\bm \beta))^\top (\bm Z_i \bm \Gamma\bm Z_i^\top + \bm \Lambda_i )^{-1}(\bm y_i - \bm f_i(\bm \beta)) 
 + \frac{1}{2}\ln|\bm Z_i \bm \Gamma\bm Z_i^\top + \bm \Lambda_i  |
 \right)
\]
Here, $\bm y_i \in \mathbb{R}^{n_i}$, where $n_i$ is the number of observations in the $i$th group. 
To trim observations within the group, we introduce auxiliary variables $\bm w_i \in \mathbb{R}^{n_i}$, and define 
\[
\begin{aligned}
\bm r_i := \bm y_i - \bm f_i (\bm \beta), \quad \bm W_i := \mbox{diag}(\bm w_i), \quad \sqrt{\bm W_i} := \mbox{diag}(\sqrt{\bm w_i}).
\end{aligned}
\]
We now form the objective 
\begin{equation}
\label{eq:trimGroup}
\begin{aligned}
\frac{1}{2} \bm r_i^\top \sqrt{\bm W_i}\left(\sqrt{\bm W_i} \bm Z_i\bm \Gamma\bm Z_i^\top\sqrt{\bm W_i} + \bm \Lambda_i^{\odot \bm w_i} \right)^{-1}\sqrt{\bm W_i} \bm r_i + 
\frac{1}{2}\ln\left|\sqrt{\bm W_i} \bm Z_i \bm \Gamma\bm Z_i^\top \sqrt{\bm W_i} + \bm \Lambda_i^{\odot \bm w_i}\right|,
\end{aligned}
\end{equation}
where $^\odot$ denotes the elementwise power operation:
\begin{equation}
\label{eq:notationDef}
 \bm \Lambda_i^{\odot \bm w_i} := 
\begin{bmatrix}
(\lambda_{i\red{1}})^{w_{i1}} & 0 & \dots &  0 \\
0 & \ddots & \ddots& \vdots \\
0 & \dots & 0 & (\lambda_{in_i})^{w_{in_i}}. 
\end{bmatrix}
\end{equation}
When $w_{ij} =1$, we recover  the  contribution of the $ij$th observation to the original likelihood. As $w_{ij} \downarrow  0$,
The $ij$th contribution to the residual is correctly eliminated by $\sqrt{w_{ij}} \downarrow 0$. 
The $j$th row and column of $\sqrt{\bm W_i} \bm Z_i \bm \Gamma\bm Z_i^\top \sqrt{\bm W_i}$ both  
go to $0$, while the $j$th entry of $\bm \Lambda_i^{\odot \bm w_i}$ goes to $1$, which removes all 
impact of the $j$th point. \red{Specifically, the matrix $\sqrt{\bm W_i} \bm Z_i \bm \Gamma\bm Z_i^\top \sqrt{\bm W_i} + \bm \Lambda_i^{\odot \bm w_i}$ with $w_{ij}=0$ and all remaining $w_{i\cdot} = 1$ is the same as the  matrix $\bm Z_i \bm \Gamma\bm Z_i^\top + \bm \Lambda_i$ obtained after deleting the $ij$-th point.}


Combining trimmed ML with priors and constraints, we obtain the following \red{modified log-likelihood}: 
\begin{equation}
\label{eq:trimGroupFull}
\begin{aligned}
\min_{\blue{\bm \theta}, \bm w}  \mathcal{L}(\blue{\bm\theta}, \bm w) := & 
\sum_{i=1}^m
\frac{1}{2} \bm r_i^\top \sqrt{\bm W_i}
\left(\sqrt{\bm W_i} \bm Z_i \bm \Gamma\bm Z_i^\top\sqrt{\bm W_i} + \bm \Lambda_i^{\odot \bm w_i} \right)^{-1}
\sqrt{\bm W_i} \bm r_i + \\
&\frac{1}{2}\ln\left|\sqrt{\bm W_i} \bm Z_i \bm \Gamma\bm Z_i^\top \sqrt{\bm W_i} + \bm \Lambda_i^{\odot \bm w_i}\right|
+ \rho(\bm \beta, \bm \gamma, \bm \Lambda)\\
& \mbox{s.t.} \quad \bm r_i = \bm y_i - \bm f_i (\bm \beta), \quad \bm 1^\top \bm w = h, \quad \bm 0 \leq \bm w \leq \bm 1,  
\quad \bm C\left(\begin{matrix} \bm \beta\\ \bm \gamma\\ \bm \Lambda \end{matrix}\right) \leq \bm c.
\end{aligned}
\end{equation}

\red{Problem}~\eqref{eq:trimGroupFull} has not been previously considered in the literature. 
We present a specialized algorithm \red{and analysis} in the next section. 

\subsection{Fitting Trimmed Constrained MEs: Algorithm and Analysis}
\label{sec:algo}

\red{Problem}~\eqref{eq:trimGroupFull} is nonsmooth and nonconvex. The key to algorithm design and analysis is to 
decouple this structure, and reduce the estimator to solving a smooth nonconvex value function over a convex set. This allows 
an efficient approach that combines classic nonlinear programming with first-order approaches for optimizing nonsmooth nonconvex problems. 
We partially minimize with respect to $(\bm \beta, \bm \gamma, \bm \Lambda)$ using an interior point method, and then 
optimize the resulting value function with respect to $\bm w$ using a first-order method.
The approach leverages ideas from variable projection~\citep{golub1973differentiation,golub2003separable,aravkin2012estimating,aravkin2018efficient}. 



We define $\bm \theta = (\bm \beta, \bm \gamma, \bm \Lambda)$,  
the implicit solution $\bm \theta(\bm w)$ and value function $v(\bm w)$ as follows:
\begin{equation}
\label{eq:v}
\begin{aligned}
\bm \theta(\bm w) & := \arg\min_{\bm \theta} \mathcal{L}(\bm \theta, \bm w) \quad \mbox{s.t.} \quad \bm C(\bm \theta) \leq \bm c\\
v(\bm w) &:=  \min_{\bm\theta} \mathcal{L}(\bm \theta, \bm w) \quad \mbox{s.t.} \quad \bm C(\bm \theta) \leq \bm c
\end{aligned}
\end{equation}
where $\mathcal{L}(\bm\theta, \bm w)$ is given in~\eqref{eq:trimGroupFull}. \red{The term $\bm\theta(\bm w)$ refers to the entire set of minimizers for a given $w$, 
which may not necessarily be a singleton.}

\blue{We first develop conditions and theory to guarantee the existence of minimizers $\bm \theta(\bm w)$. 
While there are  results in the literature for particular classes of linear mixed effects models, conditions for the existence of minimizers
for the partial nonlinear case~\eqref{eq:trimGroupFull} have not been derived. Both~\cite{harville2018linear} and~\cite{davidian1995nonlinear}
analyze the Aitken model, where $\red{\bm \Lambda} = \sigma^2 \bm H$ for a nonsingular $\bm H$, by essentially 
deriving the closed form estimates for $\bm\theta$ in this case, where the conditions that the residual is not exactly $0$ guarantees a positive estimator for $\sigma^2$.
In the nonlinear case~\eqref{eq:trimGroupFull}, this is not possible, and to guarantee existence of minimizers we have to obtain some conditions 
for model validity. \\
Let $\mathcal{S}_{++}$ denote the \red{set} of positive definite matrices, 
and for $\bm M \in \mathcal{S}_{++}$ consider the function 
\begin{equation}
\label{eq:fmr}
f(\bm M, \bm r) = \bm r^T \bm M^{-1} \bm r + \ln|\bm M|.
\end{equation}
To connect the general functional form~\eqref{eq:fmr} with the problem~\eqref{eq:trimGroupFull}, we specify functional 
domains for feasible $\bm r$ amd $\bm M$. 
\begin{definition}[Domains]
We define domains $\mathcal{D}_r$ for $\bm r$ and $\mathcal{D}_M$ for $\bm M$ as follows: 
\[\begin{aligned}
 \mathcal{D}_r &: = \{\bm y - \bm f(\bm \beta): \forall \bm \beta\}, \quad\\
\mathcal{D}_M &: = \{ \bm Z \mathrm{Diag}(\bm \gamma) \bm Z^\top + \bm\Lambda(\bm \sigma): \forall \bm \gamma \ge 0, \forall \bm \sigma> 0\}.
\end{aligned}\]
\end{definition}
Using these definitions, we can write our objective as 
\begin{equation}
  \label{eq: simple_obj}
  \min_{\bm r  \in\mathcal{D}_r, \bm M \in\mathcal{D}_M} f(\bm r, \bm M)
\end{equation}
for $f$ in~\eqref{eq:fmr}.
We also need to  define the {\bf level set} of a function. 
\begin{definition}[Level set]
The $\alpha$-level set of a function $f:\mathbb{R}^k \rightarrow \mathbb{R}$, denoted $\mathcal{L}_{f,\alpha}$, is defined by 
\[
\mathcal{L}_{f,\alpha} := \{\bm \theta: f(\bm \theta) \leq \alpha\}
\]
\end{definition}
To guarantee the existence of estimators $(\bm \theta)$ we make two assumptions. 
\begin{assumption}
\label{ass: Dr_closed}
We assume that the image of $\bm f$ is closed. This implies that $\mathcal{D}_r$ is closed.
\end{assumption}
\begin{assumption}
\label{ass: r_nonzero}
Denote the eigenvalue decomposition for  any $\bm M \in \mathcal{D}_m$ by $\bm M = \bm X_M \mbox{Diag}(\bm \lambda_M) \bm X_M^\top$.
For $\forall \bm r \in \mathcal{D}_r$ and $\forall \bm M \in \mathcal{D}_M$, 
we assume there exist $\alpha > 0$ such that 
\begin{equation}
\label{eq:keyeq}
\min_{i} \max(|\tilde r_i|, \lambda_i) \ge \alpha
\end{equation}
where $\tilde{\bm r} = X_M^\top \bm r$ and $\lambda_i$ is the corresponding eigenvalue.
\end{assumption}
Assumption~\ref{ass: r_nonzero} tells us that, after being pre-whitened by the covariance matrix, either the residual is bounded away from 0 or else the corresponding eigenvalue
is bounded away from zero. These assumptions are necessary and sufficient for the existence of minimizers. 
Before we proceed, we make a simple remark about the meta-analysis case. 
\begin{remark}
\label{rem:meta}
Assumption~\ref{ass: r_nonzero} is always satisfied for the case of meta-analysis, as long as all reported covariance matrices are positive definite. 
\end{remark} 
The remark follows because in the meta-analysis case, $\bm\Lambda$ is block diagonal,  with blocks precisely the covariance matrices reported by studies. 
For many functions $\bf f$, Assumption~\ref{ass: Dr_closed} is satisfied, including all linear and piecewise linear-quadratic functions. 
Functions that may violate these assumptions include logarithms and fractional functions, and complex models may therefore require further analysis.  We can now prove the following lemma. 
\begin{theorem}
\label{lem: level_set_compact}
Under Assumption~\ref{ass: Dr_closed} and Assumption~\ref{ass: r_nonzero}, all level sets $\mathcal{L}_{f,\alpha}$ for $f$ in \eqref{eq: simple_obj} are bounded.
Moreover, all level sets $\mathcal{L}_{f,\alpha}$ for $f$ in \eqref{eq: simple_obj} are closed.
\end{theorem} 
{\bf Proof:}
We first prove that 
\[
\|\bm r\|_2^2 + \|\bm M\|_F^2 \rightarrow \infty \quad \Rightarrow \quad f(\bm r, \bm M) \rightarrow \infty,
\]
which implies bounded level sets.  From \eqref{eq: simple_obj} and the eigenvalue decomposition, we have
\[
\begin{aligned}
f(\bm r, \bm M) =& \frac{1}{2}\sum_{i=1}^m\left[\frac{\tilde{r}_i^2}{\lambda_i} + \ln(\lambda_i)
\right]\\
\ge & \frac{1}{2}\sum_{i=1}^m \max\left(\ln(\lambda_i), 1 + \ln(\tilde{r}_i^2)\right) \ge \frac{m}{2} \min(\ln(\alpha), 1 + 2\ln(\alpha))
\end{aligned}\]
When $\|\bm r\|_2^2 + \|\bm M\|_F^2 = \|\tilde{\bm r}\|_2^2 + \|\bm \lambda_M\|_2^2 > c$, we know that there at least exists one $\tilde r_i$ or $\lambda_i$
such that $\tilde r_i^2 > c/(2m)$ or $\lambda_i^2 > c/(2m)$, and
\begin{equation}
\label{eq:lower-bound}
f(\bm r, \bm M) \ge \frac{1}{2}\max\left(\ln\left(\sqrt{\frac{c}{2m}}\right), 1 + \ln\left(\frac{c}{2m}\right)\right) + \frac{m-1}{2}\min(\ln(\alpha), 1 + 2\ln(\alpha))
\end{equation}
And we know when $c \rightarrow \infty$, $f(\bm r, \bm M) \rightarrow \infty$. \\
Next we prove that $\mathcal{L}_{f,\alpha}$ are also closed. 
Since $f$ is the continuous function and $\mathcal{D}_r$ is closed from Assumption~\ref{ass: Dr_closed}, we only need to
show that the intersection between any $\mathcal{L}_c$ and open boundary of $\mathcal{D}_M$, denoted as $\mathcal{B}_M$, is empty.
And we know that $\mathcal{B}_M$ is contained in the space of positive indefinite matrices. 
For any $\overline{\bm M} \in \mathcal{B}_M$, and any sequence ${\bm M^i} \subset \mathcal{D}_M$ approaching $\overline{\bm M}$, 
without loss of generality, we assume $\overline{\lambda_m} = 0$, and we know that, $\lambda^i_m \rightarrow 0$ and the corresponding $|\tilde r^i_m|$ has to be
stay above $\alpha$ (from Assumption~\ref{ass: r_nonzero}). 
And from the inequality~\eqref{eq:lower-bound}, we have
\[\begin{aligned}
f(\bm r, \bm M^i) \ge \frac{1}{2}\left(\frac{\alpha^2}{\lambda_m^i} + \ln(\lambda_m^i)\right) + \frac{m-1}{2}\min(\ln(\alpha), 1 + 2\ln(\alpha))
\end{aligned}\]
And we know as $\lambda^i_m \rightarrow 0$, $f(\bm r, \bm M^i) \rightarrow \infty$, therefore there exists $N$ such that $f(\bm r, \bm M^i) > c$ for all $i \ge N$,
and $\overline{\bm M} \not\in \mathcal{L}_c$.\\
We therefore have an immediate corollary. 
\begin{corollary}[Existence of minimizers]
Under Assumptions~\ref{ass: r_nonzero} and~\ref{ass: Dr_closed}, the set of minimizers $f$ is nonempty. 
\end{corollary}
By Theorem~\ref{lem: level_set_compact}, the level sets of $f$ are closed and bounded, hence compact. A continuous function assumes its minimum 
on a compact set, therefore the set of minimizers is nonempty. We can also use Theorem~\ref{lem: level_set_compact} to prove specific results 
about existence of minimizers for~\eqref{eq:trimGroupFull} under specific parametrizations, including meta-analysis and more standard longitudinal assumptions.  
Linear constraints will not fundamentally change the situation, so long as they allow any feasible solution, as is stated in the next corollary. 
\begin{corollary}[Effects of constraints]
As long as the intersection of domains $\mathcal{D}_r$ and $\mathcal{D}_M$ with the polyhedral set~$\bm C \bm \theta \leq \bm c$ is nonempty, 
minimizers exist under Assumptions~\ref{ass: r_nonzero} and~\ref{ass: Dr_closed}. Moreover, constraints can be used to ensure Assumption~\ref{ass: Dr_closed}, 
for specific cases, by ensuring that the intersection of the feasible region with $\mathcal{D}_r$ is closed.
\end{corollary}
This follows immediately from the fact that the intersection of a compact set with any closed set remains compact. 
}

\blue{Theorem~\ref{lem: level_set_compact} and following corollaries establish conditions for the existence of $\bm\theta(w)$, 
given assumptions~\ref{ass: r_nonzero} and~\ref{ass: Dr_closed}. These assumptions must hold for all $\bm w$ in the capped simplex 
specified by the modeler through the $h$ parameter, which roughly means that for any 
selection  of $h$ datapoints, the problem has to be well defined in the sense of assumptions~\ref{ass: r_nonzero} and~\ref{ass: Dr_closed}. 
}

\red{The existence of global minimizers $\bm \theta(\bm w)$ underpins the approach, which at a high level optimizers the value function 
$\bm v( \bm w)$ over the capped simplex to detect outliers. The optimization problem~\eqref{eq:trimGroupFull} has a nonconvex objective 
and nonlinear constraints. Nevertheless,  we can specify the conditions under which $\bm v(\bm w)$ is a differentiable function, and we can find 
an explicit formula for its derivative. The result is an application of the Implicit Function Theorem to the Karush Kuhn Tucker conditions that characterize
optimal solutions of~\eqref{eq:trimGroupFull}. We summarize the relevant portion of this classic result presented by~\cite[Theorem 4.4]{still2018lectures}, and 
refer the reader to~\cite{craven1984non,bonnans2013perturbation} where these results are extended under weaker conditions.}


\begin{theorem}[Smoothness of the value function]
\label{cor:expfit}
\red{For a given $\bm w$, consider $\bm\theta(\bm w)$ and let $\mathcal{I}_0$ denote the set of active constraints:
  \[
\mathcal{I}_0 := \{i: \bm C_i(\bm \theta(\bm w)) = \bm c_i\}.
 \]  
Consider the extended Lagrangian function restricted to this index set: 
\[
L(\bm \theta, \bm \mu, \bm w) = \mathcal{L}(\bm\theta, \bm w) +\sum_{i \in I_0} \mu_i (\bm C_i (\bm \theta) - \bm c_i)
\]
Suppose that the following conditions hold:
\begin{itemize}
\item Stationarity of the Lagrangian: there exist multiplies $\bm \mu(\bm w)$ with   
\[
\nabla_{\bm \theta} L = \nabla_{\bm \theta}  \mathcal{L}(\bm\theta(\bm w), \bm w)  + \sum_{i \in I_0} \mu_i(\bm w) \nabla_{\bm \theta} \bm C_i (\bm \theta(\bm w)) = 0
\]
\item Linear independence constraint qualification
\[
 \nabla_{\bm \theta}\bm C_i (\bm \theta(\bm w)), i \in \mathcal{I}_0 \quad \mbox{are linearly independent}  
 \]
\item Either of the following conditions hold:
\begin{itemize}
\item Strict complementarity:  number of active constraints is equal to the number of elements in $\bm \theta$, and all $\mu_i(\bm w)>0$.
\item Second order condition: $\nabla^2_{\bm \theta} L$ is positive definite when restricted to the tangent space $\mathcal{T}$ induced by the active constraints:
\[
\mathcal{T} = \{d: \nabla_{\bm \theta} C_i(\bm w) d = 0, i \in \mathcal{I}_0.\}
\]
\end{itemize}
\end{itemize}
Then the value function $v(\bm w)$ is differentiable, with derivative given by 
\begin{equation}
\label{eq:valDeriv}
\begin{aligned} 
\nabla v(\bm w) & = \nabla_{\bm w}  L(\bm \theta,  \bm \mu, \bm w)|_{\bm \theta (\bm w),  \bm \mu (\bm w), \bm w} \\
& = \nabla_{\bm w} \mathcal{L}(\bm\theta, \bm w)|_{\bm \theta(\bm w), \bm w}. 
\end{aligned}
\end{equation}
}
\end{theorem}

\red{The second order condition above guarantees that $\bm\theta(\bm w)$ is isolated (locally unique) minimizer~\cite[Theorem 2.5]{still2018lectures}. }

Partially minimizing over $\bm \theta$ reduces  problem~\eqref{eq:trimGroupFull}  to
\begin{equation}
\label{eq:trimGroupFullVP}
\min_{\bm w} v(\bm w) \quad  \mbox{s.t.} \quad \bm 1^\top \bm w = h, \quad \bm 0 \leq \bm w \leq \bm 1,
\end{equation}
where $v(\bm w)$ is a continuously differentiable nonconvex function, and the constrained set is the (convex) capped simplex $\Delta_h$ introduced in the trimming 
section.

The high-level optimization over $\bm w$ is implemented using projected gradient descent: 
\begin{equation}
\label{eq:value}
\bm w^+ = \proj_{\Delta_h} (\bm w - \alpha \nabla v(\bm w)),
\end{equation}
\blue{where $\proj_{\Delta_h}$ is the projection operator onto a set defined in the introduction. Since $\Delta_h$  is a convex set, the projection is unique. }

Each update to $\bm w$ requires computing the gradient $\nabla v$, which in turn requires solving for $\bm \theta$; see~\eqref{eq:v}.
The explicit implementation equivalent to~\eqref{eq:value} is summarized in Algorithm~\ref{alg:pg}. 
\blue{Projected gradient descent is guaranteed to converge to a stationary point in a wide range of settings, for any semi-algebraic 
differentiable function $v$~(\cite{attouch2013convergence}).}

\begin{algorithm}[h!]
\caption{Projected gradient descent on the Value Function $v$ of~\eqref{eq:v}}
\label{alg:pg}
\begin{algorithmic}[1]
\State {\bfseries Input:} $\bm w_0, \lambda_{\bm w}.$
\State {\bfseries Initialize:} $\nu=0$ 
\While{not converged}
\Let{$\nu$}{$\nu+1$}
\Let{$\bm \theta^{\nu+1}$}{$\displaystyle\blue{\arg} \min_{\bm \theta} \mathcal{L}(\bm \theta, \bm w^\nu) \quad \mbox{s.t} \quad \bm C(\bm \theta) \leq \bm c$}
\Let{$\bm w^{\nu+1}$}{$\displaystyle\proj_{\Delta_h}(\bm w - \alpha\blue{\nabla}_{\bm w}\mathcal{L}(\blue{\bm \theta^{\nu+1},\bm w ))} $}
\EndWhile
\State {\bfseries Output:} $\bm w_\nu, \bm \theta_\nu$
\end{algorithmic}
\end{algorithm}

Step 5 of Algorithm~\ref{alg:pg} requires solving the constrained likelihood problem~\eqref{eq:trimGroupFull} with $\bm w$ held fixed. 
We solve this problem using \texttt{IPopt}~\citep{wachter2006implementation}, a robust interior point solver that allows both simple box 
and functional constraints.  
While one could solve the entire problem using \texttt{IPopt}, \blue{optimizing with simultaneously in  $\bm\theta$ and $\bm w$ turns the problem into a high dimensional
nonsmooth nonconvex problem. Instead, we treat them differently to exploit problem structure.} Typically $\bm\theta$ is small compared to $\bm w$, which is the size of the data. 
On the other hand the constrained likelihood problem 
in $\bm \theta$ is difficult while constrained value function optimization over $\bm w$ can be solved with projected gradient.  \blue{We therefore iteratively solve 
the difficult small-dimensional problem using  \texttt{IPopt}, while handling the optimization over the larger $\bm w$ variable through value function optimization 
over the capped simplex, a convex set.}

\subsection{Nonlinear Relationships using Constrained Splines}
\label{sec:spline}

In this section we discuss using spline models to capture nonlinear relationships. 
The relationships most interesting to us are dose-response relationships, which allow us to analyze 
adverse effects of risk factor exposure (e.g. smoking, BMI, dietary consumption) on health outcomes. 
For an in-depth look at splines and spline regression see~\cite{de1978practical} and~\cite{friedman1991multivariate}. 

Constraints can be used to capture expert knowledge on the shape of
such risk curves, particularly in segments informed by sparse data. 
\blue{Basic spline functionality is available in many tools and packages. The two main innovations here 
are (1) use of constraints to capture the shape of the relationship, explained in Section~\ref{sec:shape},
and (2) nonlinear functions of splines, such as ratios or differences of logs, developed 
in Section~\ref{sec:nonlin}. First we introduce basic concepts and notation for spline models.}

\subsubsection{B-splines and bases}
\label{sec:spline_basics}

\begin{figure}[h]
\centering
\includegraphics[width=0.99\textwidth]{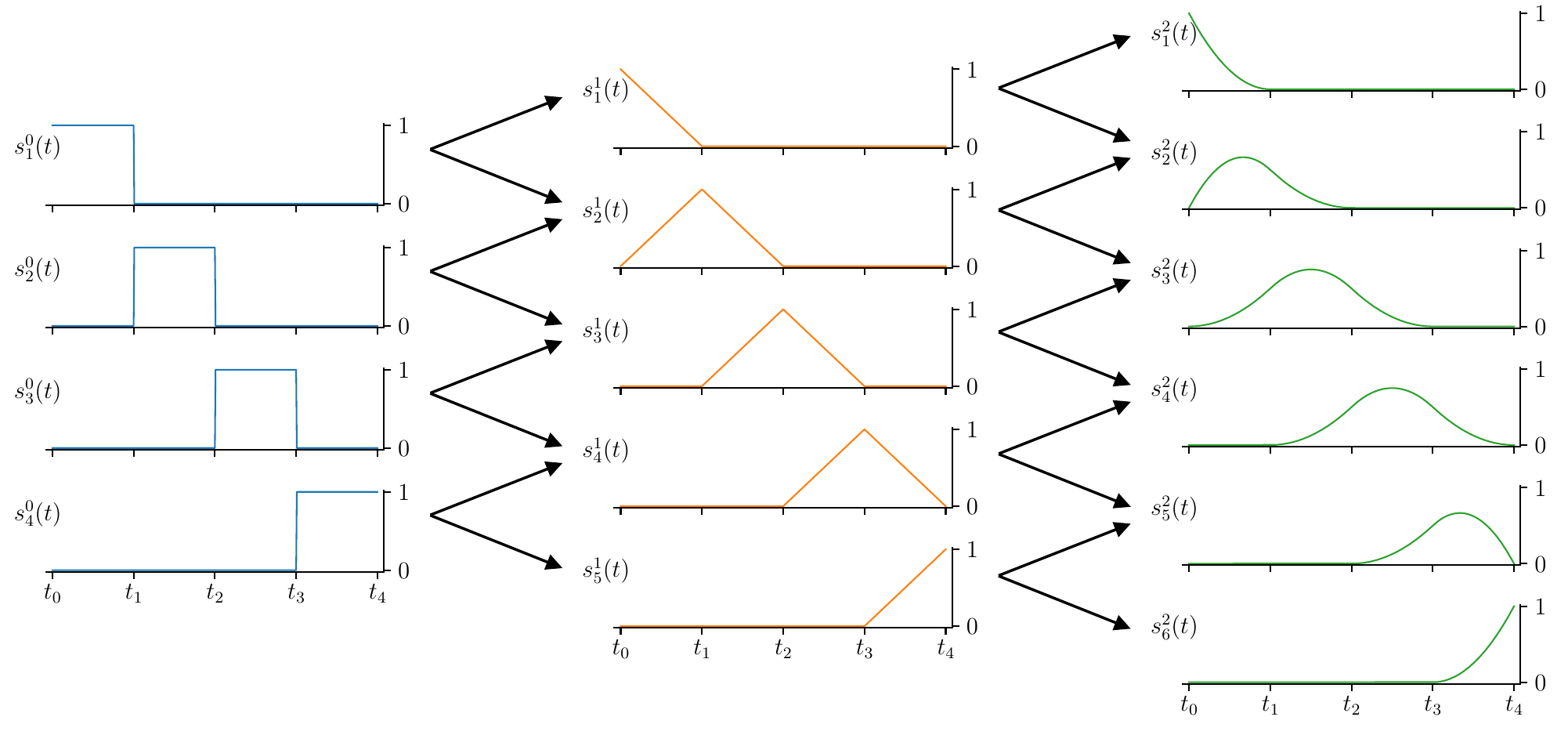}
\caption{Recursive generation of bspline basis elements (orders 0, 1, 2). \blue{The points $t_i$ in the figure are locations for knots, specified (in this case) to be equidistant. Once 
the spline bases are formed, they can then be evaluated at any points $t$ in the domain, forming $\bm X$ as in~\eqref{eq:splinedesign}.}}
\label{fig:bspline}
\end{figure}

A spline basis is a set of piecewise polynomial functions with designated degree and domain.
If we denote polynomial order by $p$, and the number of knots by $k$, we need $p+k$ basis 
elements $s^p_{j}$, which can be generated recursively  
%
%
%
as illustrated in Figure~\ref{fig:bspline}.

Given such a basis, we can represent any nonlinear curve as the linear combination of the spline basis elements, 
with coefficients $\bm \beta \in \mathbb{R}^{p+k}$: 
\begin{equation}
\label{eq:spline}
f(t) = \sum_{j=1}^{p+k} \beta_j^p s_j^p (t).
\end{equation}

These coefficients are inferred by \LimeTr analysis. 
A more standard explicit representation of~\eqref{eq:spline} is obtained by building a design matrix $\bm X$.
Given a set of \blue{$n_i$ range values $(t_1, \dots t_{n_i})$ from study $i$,}  the $j$th column of  $\bm X_i \blue{\in \mathbb{R}^{n_i \times (p+k)}}$ 
is given by the expression   
\begin{equation}
\label{eq:splinedesign}
\blue{\bm X_{i}(\cdot, j)} = \begin{bmatrix} s_j^p(t_1) \\ \vdots \\ s_j^p(\blue{t_{n_i}})\end{bmatrix}.
\end{equation}
The model for observed data coming from~\eqref{eq:spline} can now be written compactly as 
\[
\blue{\bm y_i} = \bm X_i \bm \beta +  \bm Z_i \bm u_i +\bm  \epsilon_{i},
\]
which is a special case of the main problem class~\eqref{eq:coords}.

\subsubsection{Shape constraints}
\label{sec:shape}
We can impose shape constraints such as monotonicity, concavity, and convexity on splines. 
This approach was proposed by~\cite{pya2015shape}, who used re-formulations using exponential representations of parameters to capture non-negativity. The development in this section uses explicit constraints, which is simple to encode and extends to more general cases, including functional inequality constraints $\bm C(\bm \theta) \leq \bm c$.

\paragraph{Monotonicity.}
Spline monotonicity across the domain of interest follows from monotonicity of the spline coefficients~\citep{de1978practical}. 
Given spline coefficients $\bm \beta$, 
the curve $f(t)$ in~\eqref{eq:spline} is monotonically nondecreasing when 
\[
\beta_1 \leq \beta_2 \leq \dots \leq \beta_{\blue{p+k}}
\]
and {monotonically non-increasing} if 
\[
\beta_1 \geq \beta_2 \geq \dots \geq \beta_{\blue{p+k}}.
\]
The relationship $\beta_1 \leq \beta_2$ can be written as $\beta_1 - \beta_2 \leq 0$. Stacking these inequality 
constraints for each pair $(\beta_i, \beta_{i+1})$ we can write all constraints simultaneously as 
\[
\underbrace{\begin{bmatrix}
1 & -1 & 0 & \dots & 0 \\
0 & 1 & -1 & \dots &  0 \\
\ddots & \ddots & \ddots &\ddots & \vdots\\
0 & \dots &\dots & 1 & -1
\end{bmatrix}}_{\bm C}
\begin{bmatrix} \beta_1 \\ \beta_2\\ \beta_3 \\ \vdots \\ \beta_{\blue{p+k}} \end{bmatrix} 
\leq 
\begin{bmatrix} 0 \\ 0\\ \vdots \\ 0\end{bmatrix}. 
\]
These linear constraints are a special case of the general estimator~\eqref{eq:trimGroupFull}.

\paragraph{Convexity and Concavity}

For any twice continuously differentiable function $f: \mathbb{R} \rightarrow \mathbb{R}$, convexity and concavity are captured by the 
signs of the second derivative. Specifically, $f$ is convex if $f''(t) \geq 0$ is everywhere, and concave if $f''(t) \leq 0$ everywhere. 
We can compute $f''(t)$ for each interval, and impose linear inequality constraints on these expressions. 
We can therefore easily pick any of the shape combinations given in~\cite[Table 1]{pya2015shape},
as well as imposing any other constraints on $\bm \beta$ (including bounds) through the interface of \LimeTr .

\subsubsection{Nonlinear measurements}
\label{sec:nonlin}

Some of the studies in Section~\ref{sec:real} use nonlinear observation mechanisms.
In particular, given a dose-response curve \blue{of the form~\eqref{eq:spline}},
 studies often report odds ratio of an outcome between exposed and unexposed groups 
that are defined across two intervals on the underlying curve: 
\begin{equation}
    \label{eq:indirect}
    \begin{aligned}
    \blue{\mbox{relative risk}} &= \frac{\frac{1}{a_1-a_0}\int_{a_0}^{a_1}f(t) dt}{\frac{1}{b_1- b_0}\int_{b_0}^{b_1}f(t)dt} 
    \end{aligned}
\end{equation}
When $f(t)$ is represented using a spline, each integral is a linear function of $\bm \beta$. 
If we take the observations to be the log of the relative risk \blue{for observation $j$ in study $i$}, this is given by
\[
y_{ij} := \blue{\log \mbox{relative risk for $ij$ observation}}= \ln(\langle \bm x_{ij}^1, \bm \beta \rangle) - \ln(\langle \bm x_{ij}^2, \bm \beta\rangle):= f_{ij}(\bm \beta),
\]
a particularly useful example of the general nonlinear term $\bm f_{ij}(\bm \beta)$ in problem class~\eqref{eq:coords}.
\blue{A set of examples of epidemiological models that arise from relative risk are discussed in Section~\ref{sec:smoking}.}

\subsection{Uncertainty Estimation}
\label{sec:var}

The \LimeTr  package modifies the parametric  bootstrap~\citep{efron1994introduction} to estimate the uncertainty of 
the fitting procedure. This strategy is necessary when constraints are present, and standard Fisher-based strategies for 
posterior variance selection do not apply~\citep{cox2005delta}.

The modified parametric bootstrap is similar to the standard bootstrap, but can be used more effectively for sparse data, 
e.g. when different studies sample sparsely across a dose-response curve. 
The approach can be used with any estimator~\eqref{eq:trimGroupFull}.

In the linear Gaussian case, the standard bootstrap is equivalent to bootstrapping empirical residuals, since 
every datapoint can be reconstructed this way. When the original data is sparse, we modify the parametric procedure to sample {\it modeled} residuals. Having obtained the estimate $(\bm{\hat \beta}, \bm{\hat \Lambda}, \bm{\hat \gamma})$, we sample model-based errors to get new bootstrap realizations $\bm{\bar y}$: 
\[
\quad \bm{\bar y}_i = \blue{\bm f_i( \bm{\hat \beta})} +  \bm Z_i \bm {\bar u}_i + \bm{\bar \epsilon}_i, 
\]
where $\bm{\bar \epsilon_i} \sim N(0, \hat \Lambda)$ and $\bm{ \bar u_i} \sim N(0, \bm \hat \gamma)$. 
These realizations have the same structure as the input data.
For each realization $\bm{\bar y}$, we then re-run the fit, 
and obtain  $N$ estimates $\{\bm{\hat \beta}, \bm{\hat \Lambda}, \bm{\hat \gamma})\}_{1:N}$. 
This set of estimates is used to approximate the variance of the fitting procedure along with any confidence bounds. The procedure can be applied in 
sparse and complex cases but depends 
on the initial fit $(\bm{\hat \beta}, \bm{\hat \Lambda}, \bm{\hat \gamma})$.
Its exact theoretical properties are outside the scope of the current paper and are a topic of ongoing research. 
\blue{We use N = 1000 in all of the numerical experiments in the next section. This significantly increases computational load, compared to 
a single fit. How to make the procedure more efficient (or develop alternatives) is another ongoing research topic.}

\section{Verifications}
\label{sec:verify}

In this section we validate \LimeTr  on synthetic and empirical datasets. 
In Section~\ref{sec:synth} we show how \LimeTr  compares to existing robust packages on simple problems \blue{that can be solved by 
other available tools}; see Table~\ref{table:novelty}. 
We focus on robustness of the estimates to outliers, which is a key technical contribution of the paper. 

In Section~\ref{sec:real} we use the advanced features of \LimeTr  to analyze multiple datasets in public health, where robustness to outliers, and information communicated through constraints and nonlinear measurements all play an important role. \blue{These examples illustrate advanced \texttt{LimeTr} functionality that is not available in other tools.
All examples in the paper are available online, along with a growing library of additional examples (see the Supplementary Materials section). }

\subsection{Validation Using Synthetic Data}
\label{sec:synth}

Here we consider two common mixed effects models. First we look at meta-analysis, where the $\bm \Lambda$ term is known while $(\bm \beta, \bm \gamma)$ are unknown. Then we look at a simple longitudinal case, where all three parameters are unknown, and $\bm \Lambda$ is modeled as 
$\sigma^2 \bm I$ with unknown scalar $\sigma^2$. 
In both of these cases, we compare 
the performance of \LimeTr against several available packages. The simulated data is the same for both examples; only the model is different. 

For the experiments, we generated $30$ synthetic datasets with $10$ observations in each of $10$ studies ($n=100$). The underlying true distribution is defined by $\beta_0 = 0$, $\beta_1 = 5$, $\gamma=6$, and 
$\sigma=4$, where $\gamma$ is the standard deviation of the between-study heterogeneity and $\sigma$ is the standard deviation of the measurement error. For the meta-analysis simulation, we assigned each observation a standard error of $4$. The domain of the covariate $x_1$ is $[0, 10]$. 
To create outliers, we randomly chose $15$ data points in the sub-domain $[6, 10]$ 
and offset them according to: $y'_i = y_i - 30 - |N(0, 80^2)|$. \blue{The setting of the simulation uses large outliers to show the robustness of the model to trimming. 
The larger the outliers, the larger the difference in performance between \texttt{LimeTr} and other tools. 
The behavior of trimming in the meta-analysis of real data  is more clear in Section~\ref{sec:real}.} 

\subsubsection{Meta-analysis.} 
\label{sec:example_meta}

We compared \LimeTr to three widely-used packages 
that have some robust functionality: 
\texttt{metafor}, \texttt{robumeta}, and \texttt{metaplus}. 
The functionality developed in these packages 
differs from that of \LimeTr .
The \texttt{metafor} and \texttt{robumeta} packages refer to robustness in the context of the sandwich variance estimator, which makes the uncertainty around predictions robust to correlation of observations within groups. 
\texttt{metaplus} uses heavy-tailed distributions to model random effects, which potentially allows one to account for outlying studies but not measurements within studies. 

\begin{figure}[h!]
\centering
\includegraphics[width=0.60\textwidth]{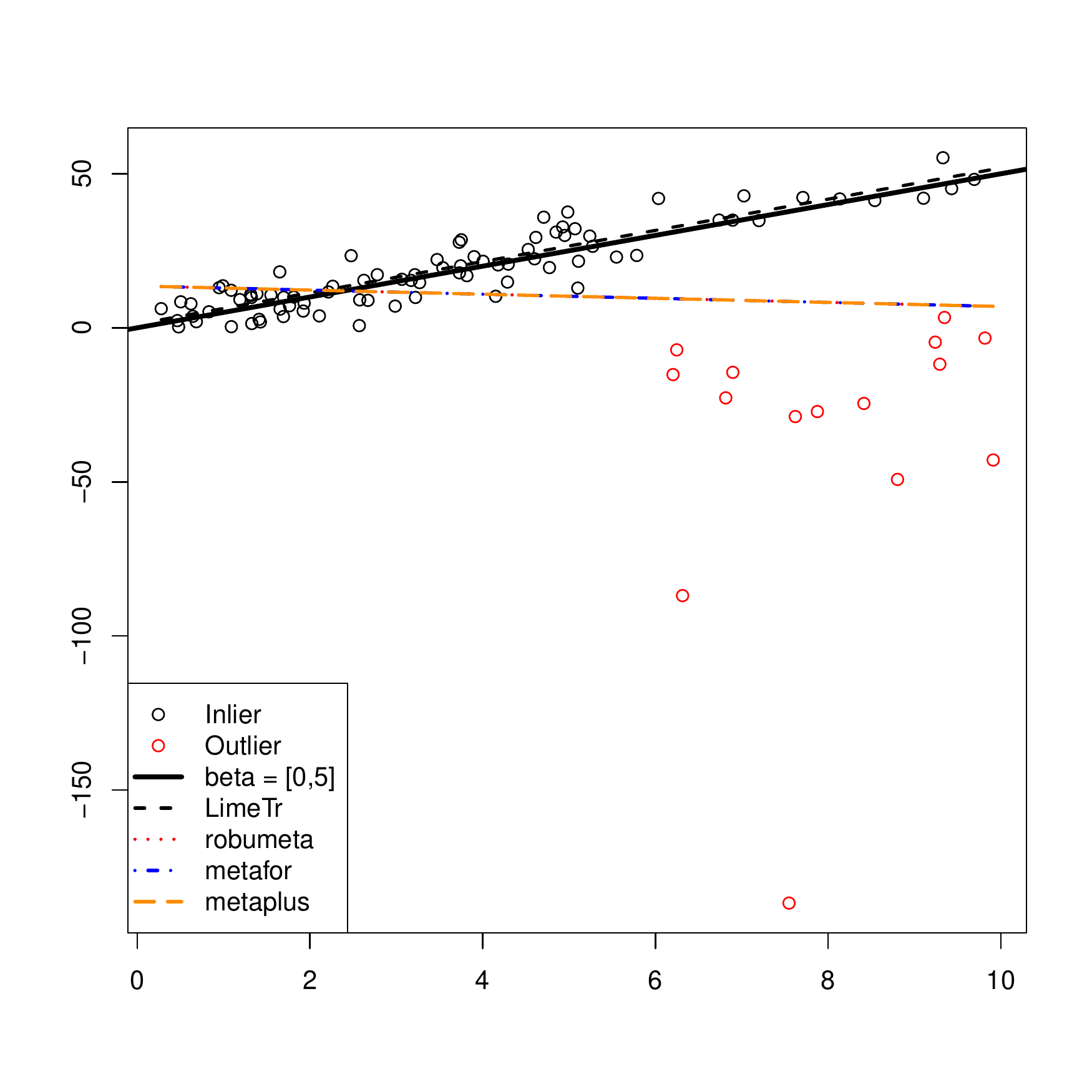}
\caption{A representative instance of the experiment summarized in Table~\ref{tab:metareg_results_case1}. 
True mechanism is shown using 
solid line; the true model is successfully inferred by the \LimeTr package despite the outliers (red points). 
\label{fig:case1_plot}}
\end{figure}

Nonetheless, it is useful to see how a new package compares
with competing alternatives on simple examples.
We compared the packages to \LimeTr  in terms of error incurred when estimating 
ground truth parameters $(\beta_0, \beta_1, \gamma)$, computation time,  
the true positive fraction (TPF) of outliers detected
and the false positive fraction (FPF) of inliers incorrectly identified. 
If a threshold of $0.8$ inliers is given to \LimeTr , 
then outliers are exactly data points with an estimated weight $w_i$ of zero, 
and those correspond to the largest absolute model residuals. 
To compare with other packages in terms of TPF and FPF,  we 
identified the 20 data points with the highest residuals according to each packages' fit.
Table~\ref{tab:metareg_results_case1} and Figure~\ref{fig:case1_plot} show the results of the meta-analysis simulation. The metrics are averages of 30 estimates from models fit on the synthetic datasets. \LimeTr  had lower absolute error in $(\beta_0, \beta_1, \gamma)$, higher TPF, lower FPF and faster computation time than the alternatives. 
\begin{table}[h!]
    \centering
    \begin{tabular}{c|c|c|c|c|c|c}\hline
 Package     & $\beta_0$& $\beta_1$ & $\red{\sqrt{\gamma}}$ & TPF & FPF & Seconds  \\\hline
      Truth &  0 & 5 & 6 & --- & --- & --- \\ \hline
      LimeTr & {\bf 0.61} & {\bf 4.95} &{\bf  4.86} & {\bf 1.00} & {\bf 0.06} & {\bf 0.35} \\ \hline
      robumeta & 10.92 & 0.03 & 37.2 & 0.68 & 0.12 & 6.54 \\ \hline
      metaplus & 10.92 & 0.03 & 35.1 & 0.68 & 0.12 & 42.4 \\ \hline\hline 
    metafor & 10.92 & 0.03 & 35.4 & 0.68 & 0.12 & 1.51 \\ \hline
    \end{tabular}
    \caption{\label{tab:metareg_results_case1}
Results of Meta-Analysis Comparison. True values are $\beta_0 = 0, \beta_1 = 5, \gamma = 6$. Results show average estimates across realizations.
    The \LimeTr package has much smaller absolute error in $\beta_0$, $\beta_1$ and in $\gamma$ than other packages.}
\end{table}

\subsubsection{Longitudinal Example.} 
\label{sec:example_long}
Here we compare \LimeTr  to R packages for fitting robust mixed effects models. Rather than assuming that errors are distributed as Gaussian, the packages use 
Huberized likelihoods (\texttt{robustlmm}) and Student's t distributions (\texttt{heavy})
to model contamination by outliers. 
\LimeTr  identifies outliers through the weights $w_i$ in the
likelihood estimate~\eqref{eq:MML} that now captures  
simple longitudinal analysis. 
Specifically, $\bm\Lambda$ is no longer specified as in the meta-analysis case, but is instead parameterized through a single unknown error $\bm\Lambda = \sigma^2 \bm I$ 
common to all observations. 
We use the same simulation structure as in Section~\ref{sec:example_meta},
now replacing observation-specific standard errors with random errors generated according to $N(0,\sigma^2)$. 
Since $\sigma^2$ is now also an unknown parameter, we 
check how well it is estimated by all packages and report this error in Table~\ref{tab:metareg_results_case2}. 
\begin{figure}[h!]
\centering
\includegraphics[width=0.60\textwidth]{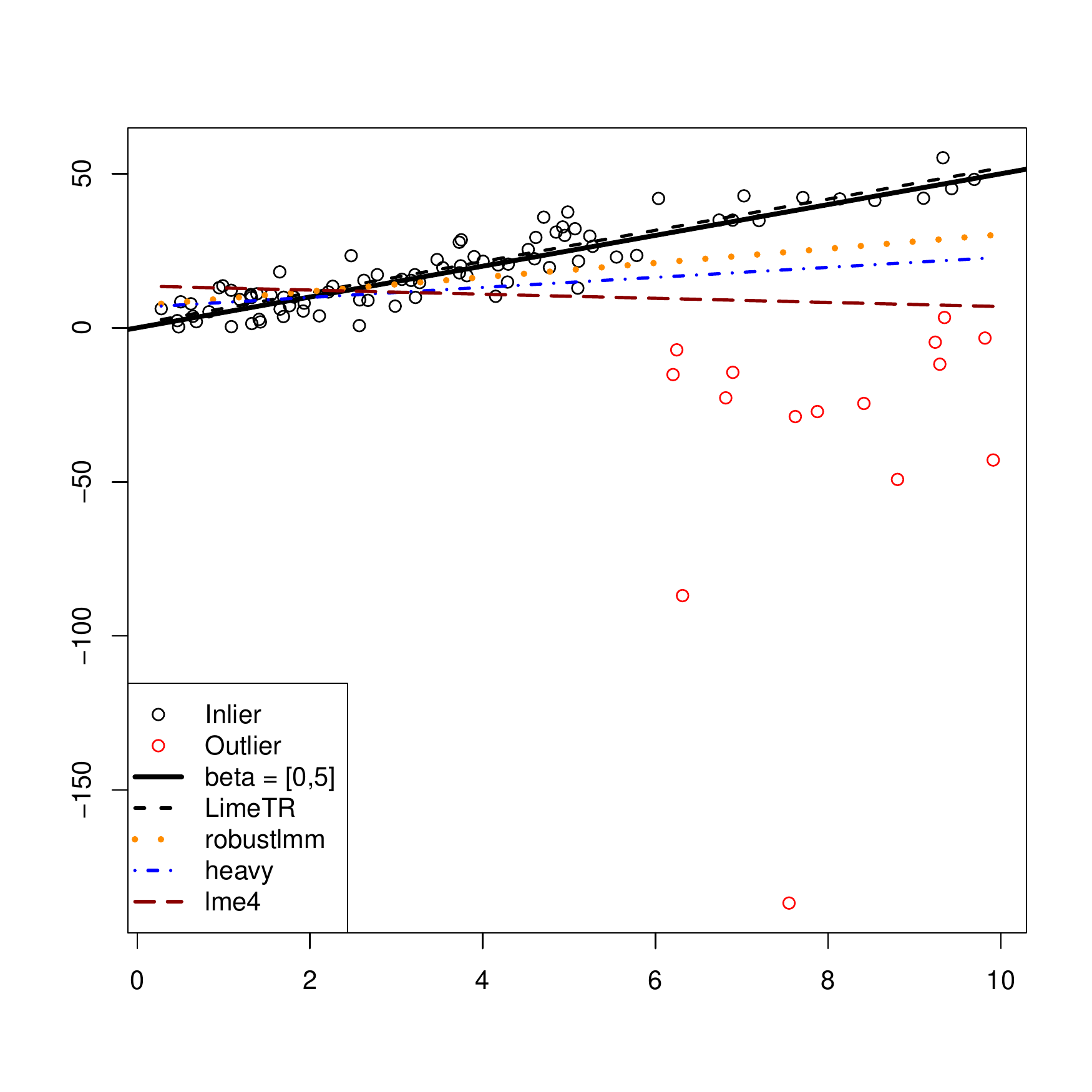}
\caption{ \label{fig:case2_plot}
A representative instance of the experiment summarized in Table~
\ref{tab:metareg_results_case2}. 
\texttt{robustlmm} and \texttt{heavy} packages both estimate $\beta$ better than \texttt{lme4}, likely because they use distributions with heavier tails.  \LimeTr outperforms all three alternatives.} 
\end{figure}
\begin{table}[h!]
    \centering
    \begin{tabular}{c|c|c|c|c|c|c|c}\hline
 Package     & $\beta_0$& $\beta_1$ & $\sqrt{\gamma}$ & $\sigma$ & TPF & FPF & Seconds  \\\hline
     Truth & 0 & 5 & 6 & 4 & --- & --- & --- \\ \hline
      LimeTr & {\bf 0.64} & {\bf 4.95} & 4.93 & {\bf 3.61} & {\bf 1.000} & {\bf 0.06} & 1.04 \\ \hline
      robustlmm & 3.32 & 3.67 & {\bf 5.14} & 9.53 & 0.99 & 0.06 & 7.7 \\ \hline
      heavy & 2.43 & 3.55 & 3.61 & NA & 0.95 & 0.07 & {0.16} \\ \hline \hline
            lme4 & 10.6 & 0.09 & {4.89} & 31.0 & 0.68 & 0.11 & {\bf 0.08} \\ \hline
    \end{tabular}
    \caption{    \label{tab:metareg_results_case2}
Results of Longitudinal Comparison. True values are $\beta_0 = 0, \beta_1 = 5, \gamma = 6, \sigma = 4$. Both \texttt{robustlmm} and the 
\texttt{heavy} package do better than the standard \texttt{lme4}. \LimeTr  more accurately estimates $\beta_0$, $\beta_1$, and $\sigma$. The 
\texttt{heavy} package 
    estimates $\gamma$ more accurately. However, estimated $\gamma$ typically increases with a worse fit to the data, and heavy does not accurately estimate $\beta$.}
\end{table}

\subsection{Real-World Case Studies}
\label{sec:real}

In this section we look at three real-data cases, all using meta-analysis. Across these examples, we show how trimming, dose-response relationships, 
and non-linear observation mechanisms come together to help understand complex and heterogeneous data. 

\subsubsection{Simple Example: Vitamin A vs. Diarrheal Disease}.
\label{sec:vitA}

\begin{figure}[h!]
\centering
\includegraphics[width=0.49\textwidth]{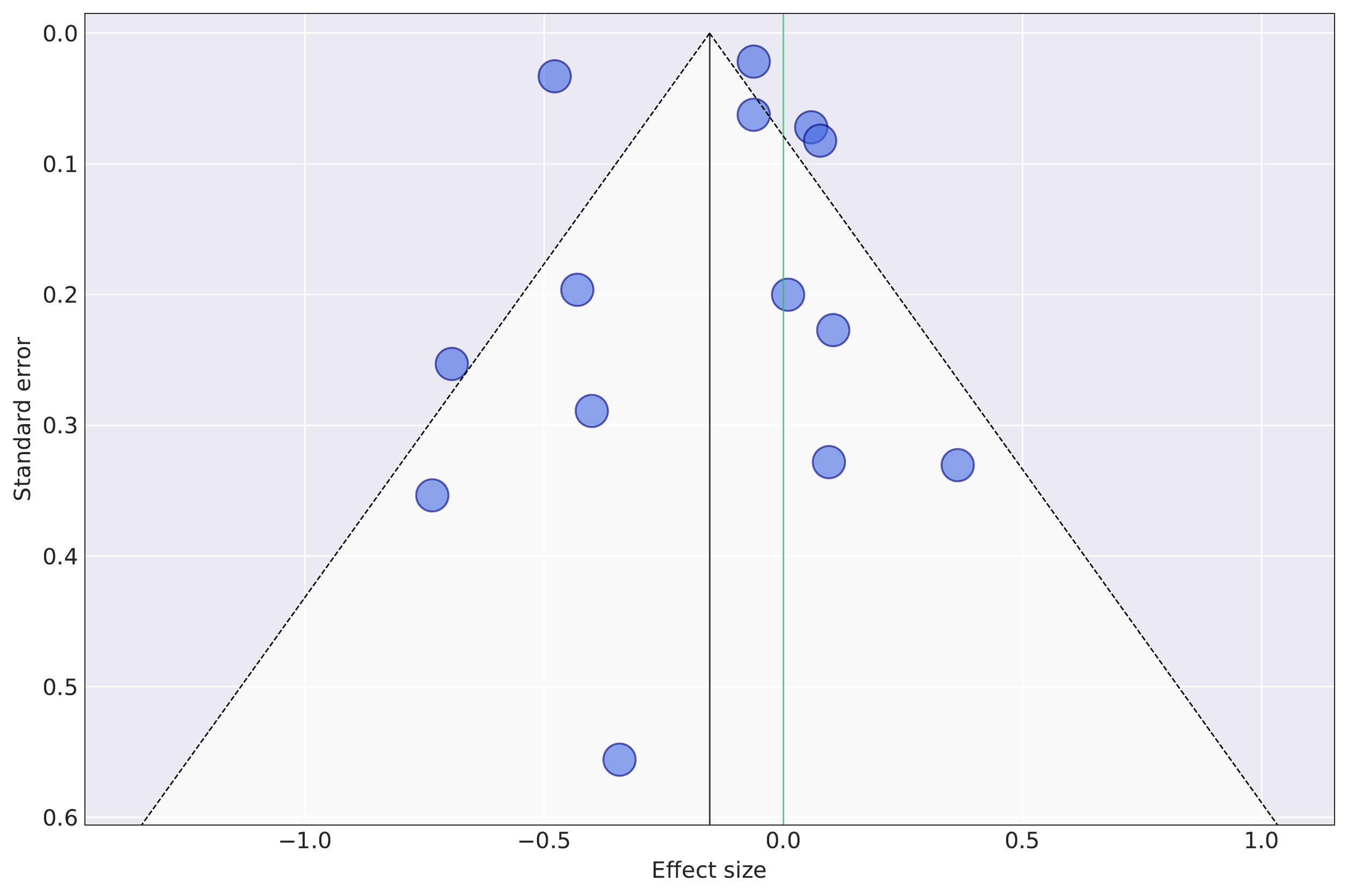}
\includegraphics[width=0.49\textwidth]{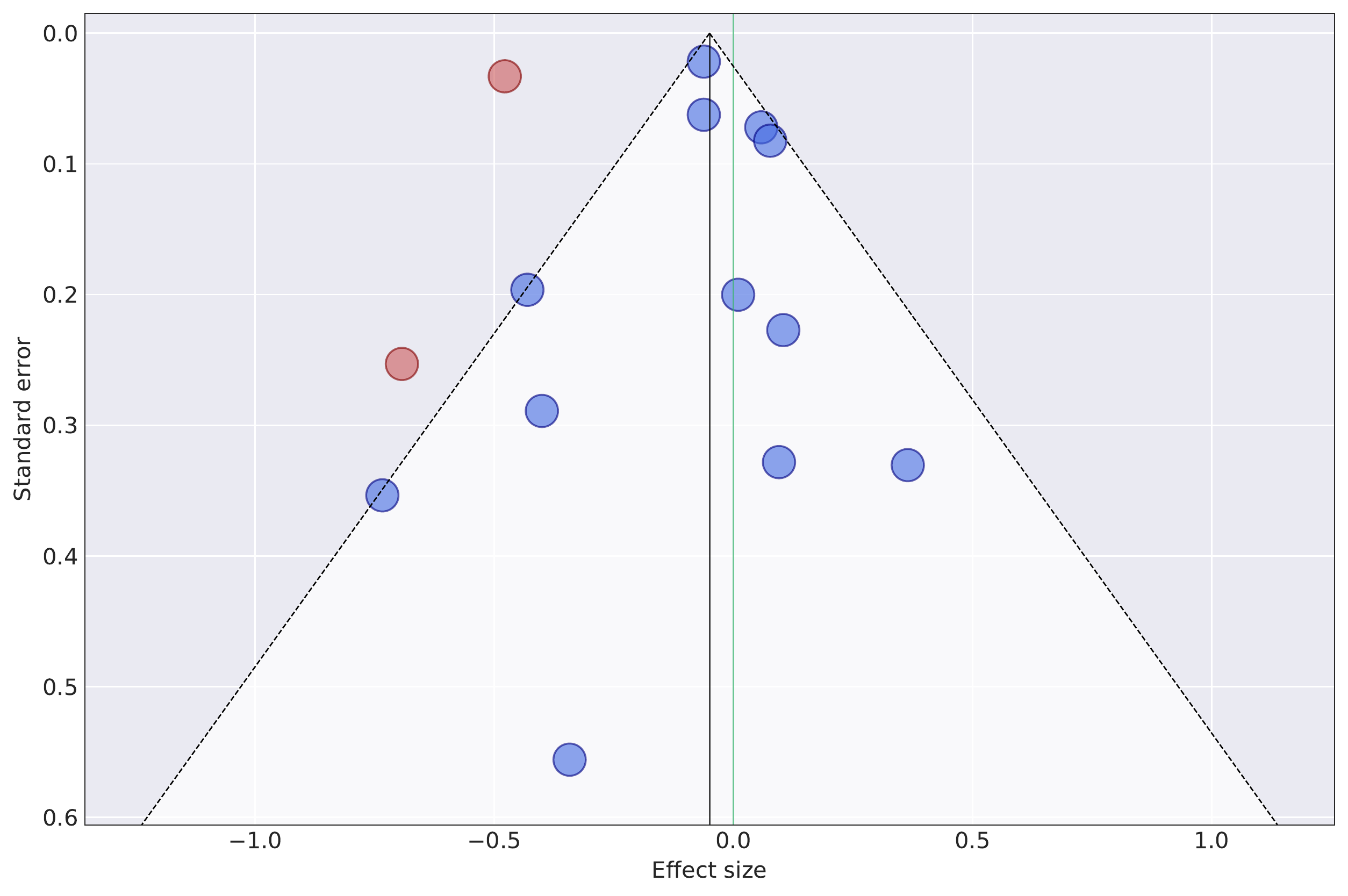}
\caption{Results for Vitamin A vs. \blue{incidence and/or mortality rates} for diarrheal disease, shown using 
the funnel plot (effect size vs. reported standard error). 
The left panel shows model without trimming, while the right panel shows the model with trimming. Trimming 10\% of the studies shifts the effect closer to $0$, making it less plausible that Vitamin A is protective,
and identifies potential outliers (right panel, shown in red), 
which are able to hide more easily in the left panel. 
\label{fig:vitA}}
\end{figure}

The first example is a simple linear model that aims to quantify the effect of vitamin A supplementation on diarrheal disease. This is an important topic in global health, and we refer the interested reader to 
the Cochrane systematic review of the topic~\citep{imdad2017vitamin}. 
In this example, we examine the influence outliers can have on inferences from a model, but we do not discuss in detail how to interpret the findings. 

In this example, the dependent variable is the natural log of relative risk of incidence \blue{or mortality from diarrheal disease from 12 studies.  
Two of the studies report on both incidence and mortality, giving 14 total datapoints}. No covariates are used.
The model we consider is  
\[
\bm y_i = \bm 1_i\beta_0  + \bm 1_i u_i + \bm \epsilon_i,
\]
where $\bm y_i$ are data sets reported from each study, $\bm 1_i$ is a vector of $1$'s of the same size 
as the number of observations for study $i$, 
$\bm \epsilon_i \sim N(\bm 0, \bm \Sigma_i)$ are associated standard errors, 
and $u_i\sim N(0, \gamma^2)$ is a study-specific random effect, with 
$\gamma^2$ accounting for between-study heterogeneity. 

Figure~\ref{fig:vitA} shows the results without trimming 
and with 10\% trimming. We included data from 12 randomized control trials in the models, with a few studies having multiple observations. Without trimming, the estimated effect size is $\beta_0 = -0.15$; with trimming it is three time smaller, $\beta_0 = -0.05$. In the trimmed model, we observe all preserved points inside the funnel, indicating that there is no expected between-study heterogeneity. 
This is confirmed by the estimates -- without trimming, between-study heterogeneity $\gamma^2$ is estimated to be 0.0435, and after trimming two studies it is reduced to $7.14 \times 10^{-9}$, nearly $0$. 
Moreover, the potential outliers are unusual: while 
all other studies deal with ages 1 year or younger, the trimmed studies are among older ages, up to 10 years in one case and 3-6 years in the other. One of the trimmed studies was also conducted in slums, a non-representative population. 

\subsubsection{Spline Example: Smoking vs. Lung Cancer.}
\label{sec:smoking}

\paragraph{log relative risk model.}
The log relative risk model is very common in the epidemiological context, and is often approximated by the log-linear \blue{relative risk} model.
\blue{A brief introduction to the nonlinear model was given in Section~\ref{sec:nonlin}. Here we derive two models, and appropriate random effect specification, 
that can be used to analyze smoking and diet data.}

\blue{In words, the log-relative risk model can be written as}
\[\begin{aligned}
\ln(\text{rel risk}) = &\left(
\frac{\ln(\text{risk at alt exposure}) - \ln(\text{risk at ref exposure})}{\text{alt exposure} - \text{ref exposure}} + \blue{\text{random effect}}
\right)\times\\
&(\text{alt exposure} - \text{ref exposure}) + \text{measurement noise}
\end{aligned}\]
where the random effect is on the average slope and  the measurement error in the log space.
Instead of \blue{making the strong assumption that  the log of the risk is a linear function of exposure, we use a spline to represent this relationship:}
\[
\ln(\text{risk at exposure } t) = \ln(\left\langle{\bm x, \bm \beta}\right\rangle)
\]
where $\bm x$ is the design vector at exposure $t$ and the spline is parametrized by $\bm \beta$, see Section~\ref{sec:spline_basics}.

The correlation between smoking and lung cancer is indisputable~\citep{gandini2008tobacco,lee2012systematic}. The exact nature of the relationship and its uncertainty requires accounting for the dose-response relationship between the amount smoked (typically measured in pack-years) and 
odds of lung cancer.  We expect a nonlinear relationship between smoking and lung cancer, and the spline methodology described in Section~\ref{sec:spline_basics} can be used. 

The outcome here is the natural log of relative risk (compared to nonsmokers). The effect of interest is a function of a continuous exposure, measured in pack-years smoked.  All studies compare different levels of exposure to non-smokers (exposure = 0), and we assume there is no risk for non-smokers ($\ln(\text{risk at exposure 0}) = 0$).
\blue{Since all datapoints share a common reference, we can simplify the relative risk model as follows:}
\[
\ln(\text{rel risk}) = \left(
\frac{\ln(\text{risk at alt exposure})}{\text{alt exposure}} +  \blue{\text{random effect}}
\right)\times \text{alt exposure} +  \blue{\text{measurement noise}}
\]
If we normalize risk of nonsmokers to $1$, and use a spline to model the nonlinear risk curve, we have the explicit expression
\[
y_{ij} = \log(\langle \bm x_{ij}, \bm \beta)) 
+ \mbox{exposure}_{ij} u_i + \epsilon_{ij}
\]
with \blue{$y_{ij}$ the log relative risk,} $\bm x_{ij}$ computed using a spline basis matrix for exposure$_{ij}$ (see Section~\ref{sec:spline_basics}), $u_i$ the random effect for study $i$, and 
$\epsilon_{ij} \sim N(\bm 0, \sigma_{ij}^2)$ the variance reported by the $i$th study for its $j$th point. 

\begin{figure}[ht!]
\centering
\includegraphics[width=0.49\textwidth]{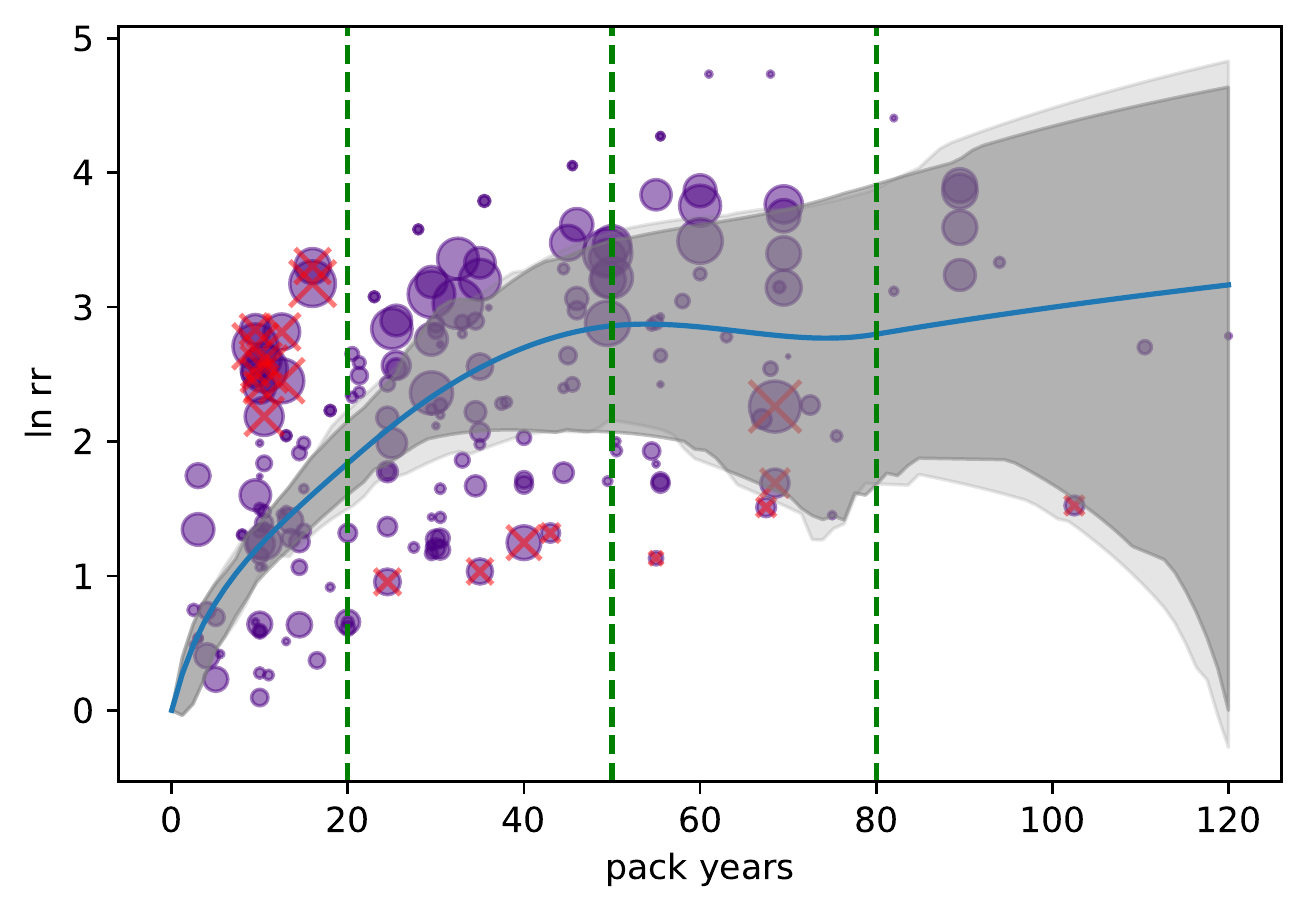}
\includegraphics[width=0.49\textwidth]{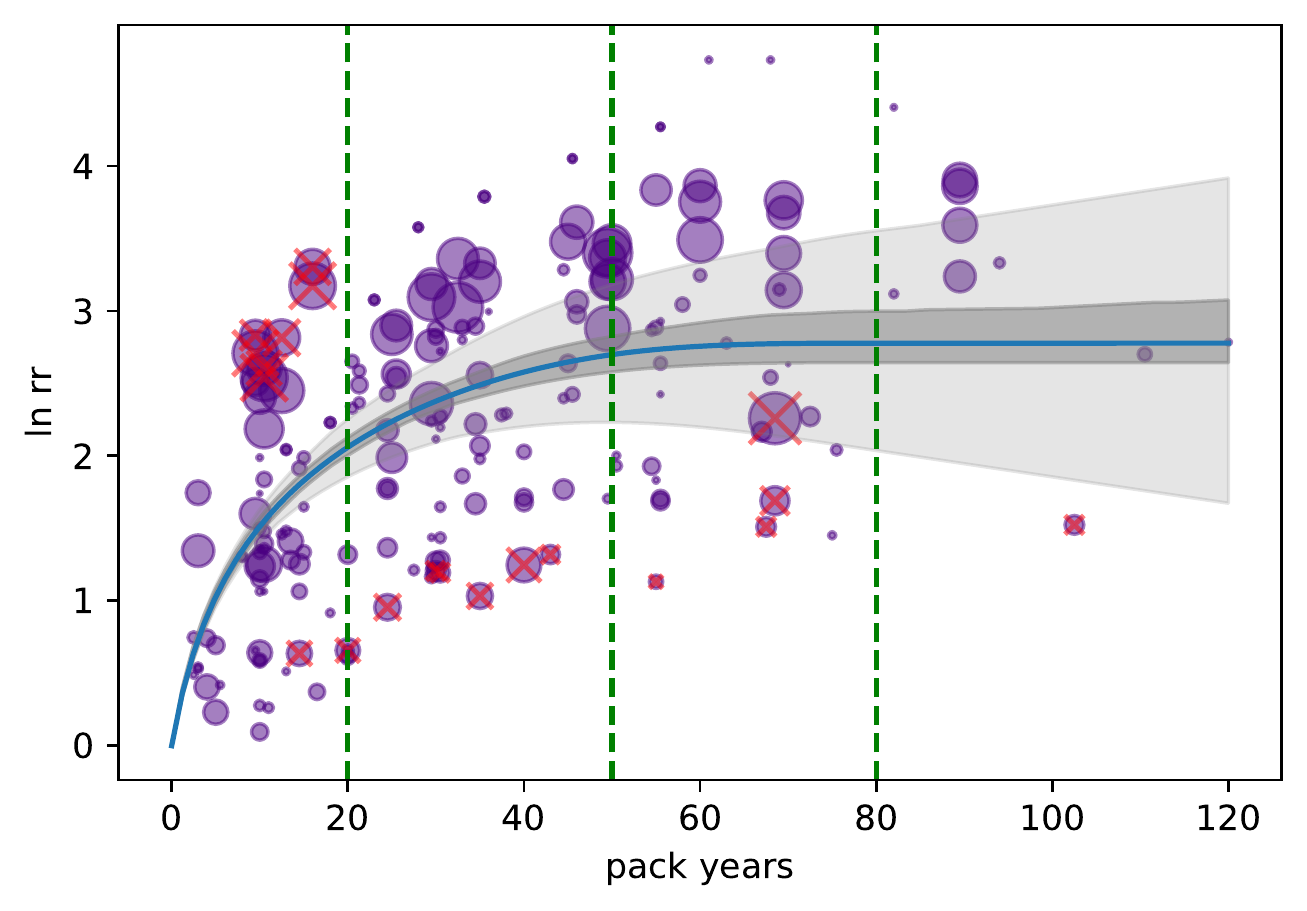}
\caption{Modeling dose-response relationship between exposure (pack years) and log-odds of lung cancer. Left figure shows a cubic spline, while right figure shows 
a cubic spline with monotonically increasing risk and concavity constraints.  \blue{The reference group is `nonsmokers' in all studies, so taking nonsmoking risk at $1$ we can 
plot the relative risk for each exposure group at its midpoint, and show the comparison to the estimated risk curve.}
The uncertainty of the mean is shown in dark grey, and additional uncertainty due to heterogeneity is shown in light gray.
Constraints regularize the shape and decrease fixed effect uncertainty, but have higher estimated heterogeneity, whereas the more flexible model explains the data and has higher fixed effect uncertainty but lower heterogeneity. 
 10\% trimming removes points that are far away from the mean dose-response relationship, as well as those moderately away from the mean but with very low reported standard deviation. Point radii on the graphs are inversely proportional to the reported standard deviations.  
\label{fig:smoking}}
\end{figure}

To obtain the results in Figure~\ref{fig:smoking} using \LimeTr , 
interior knots were set at the 10th, 50th, and 80th percentiles of 
the exposure values observed in the data,   corresponding to pack-year levels of 10, 30, and 55, respectively. 
We included 199 datapoints across 25 studies in the analysis, and again 
trimmed 10\% of the data. Trimming in this case removes datapoints that are 
far away from the group (even considering between-study heterogeneity), 
as well as points that are closer to the mean but over-confident; 
these types of outliers are specific to meta-analysis.

We also used multiple priors and constraints. First, 
we enforced that at an exposure of 0, the log relative risk must be equal to 0
(baseline: non-smokers) by adding direct constraint on $\beta_0$. 
To control changes in slope in data sparse segments, 
we included a Gaussian prior of N(0, 0.01) on the highest derivative in each segment. Additionally, the segment between the penultimate and exterior knots on the right side is particularly data sparse, and is more prone to implausible behavior due to its location at the terminus, and we force that spline segment to be linear. 

We show the unconstrained cubic spline in the left panel of Figure~\ref{fig:smoking}, and the constrained analysis that uses monotonicity and concavity of the curve in the right panel the same figure. The mean relationship looks more regular when using constraints, but the model cannot explain the data as well and so the estimate of heterogeneity is higher. On the other hand, the more flexible model has higher fixed effects uncertainty, and a lower estimate for between-study heterogeneity. 
 The point sets selected for trimming are slightly different as well between the two experiments.  

\subsubsection{Indirect nonlinear observations: red meat vs. breast cancer.}
\label{sec:meat}

The effect of red meat on various health outcomes is \red{a topic of} ongoing debate. 
In this section we briefly consider the relationship between 
breast cancer and red meat consumption, which has been systematically studied~\citep{anderson2018red}. In this section, we use data from 
available studies on breast cancer and red meat to show 
two more features of the \LimeTr  package: nonlinear 
observation mechanisms and monotonicity constraints.

The smoking example in Section~\ref{sec:smoking} uses a direct observation 
model, since all measurement are comparisons to the baseline non-smoker group. 
This is not the case for other risk-outcome pairs. When considering 
the effect of red meat consumption, studies typically report multiple comparisons
between groups that consume various amounts of meat.
In particular, all datapoints across studies are given to us a tuple: odds ratio 
for group $[a,b]$ vs. group $[c,d]$. These datapoints are thus not measuring the spline directly, but are average slopes between points in log-derivative space. 
The observation model \blue{given} by \blue{\eqref{eq:indirect}}.
\begin{figure}[ht!]
\centering
\includegraphics[width=0.49\textwidth]{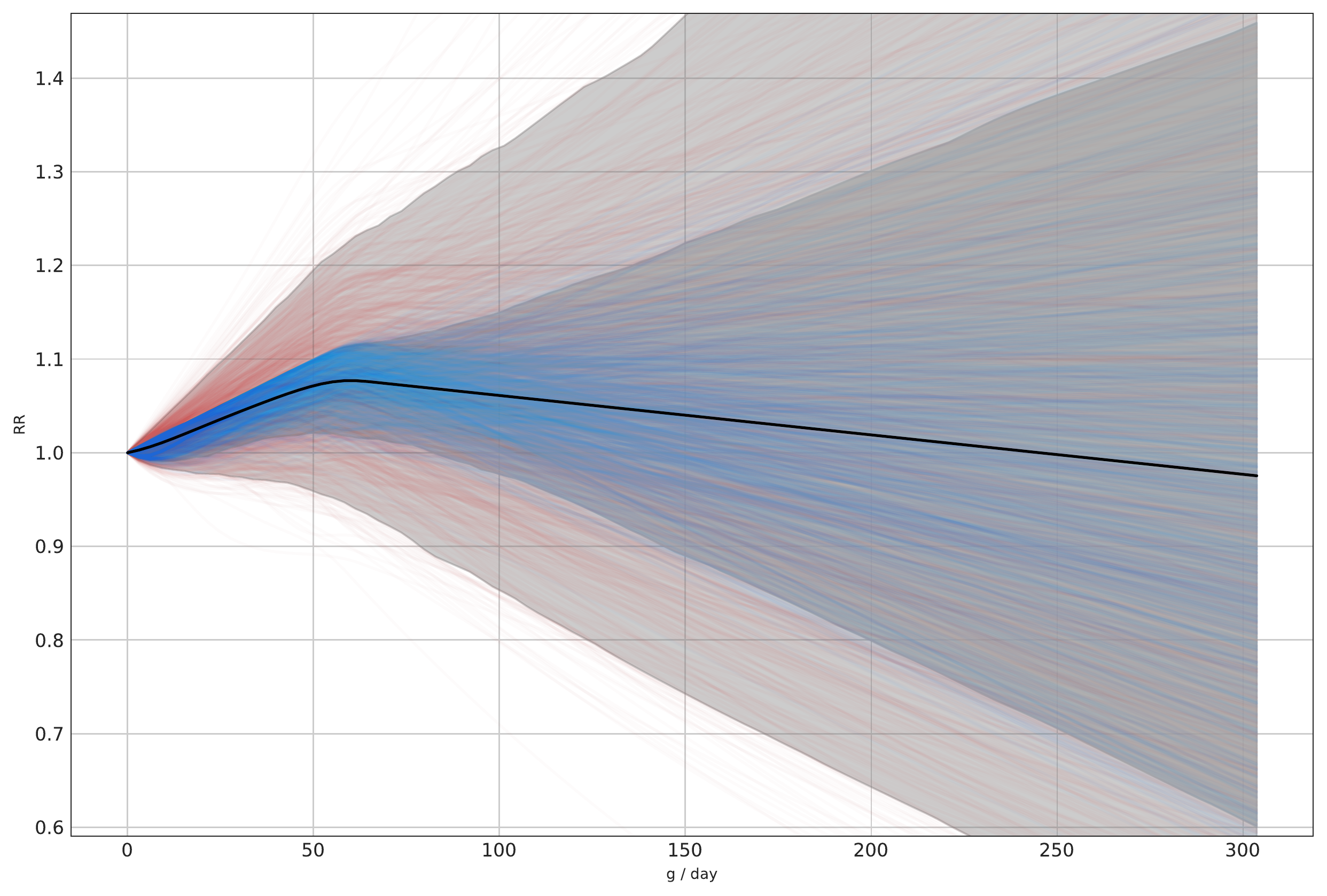}
\includegraphics[width=0.49\textwidth]{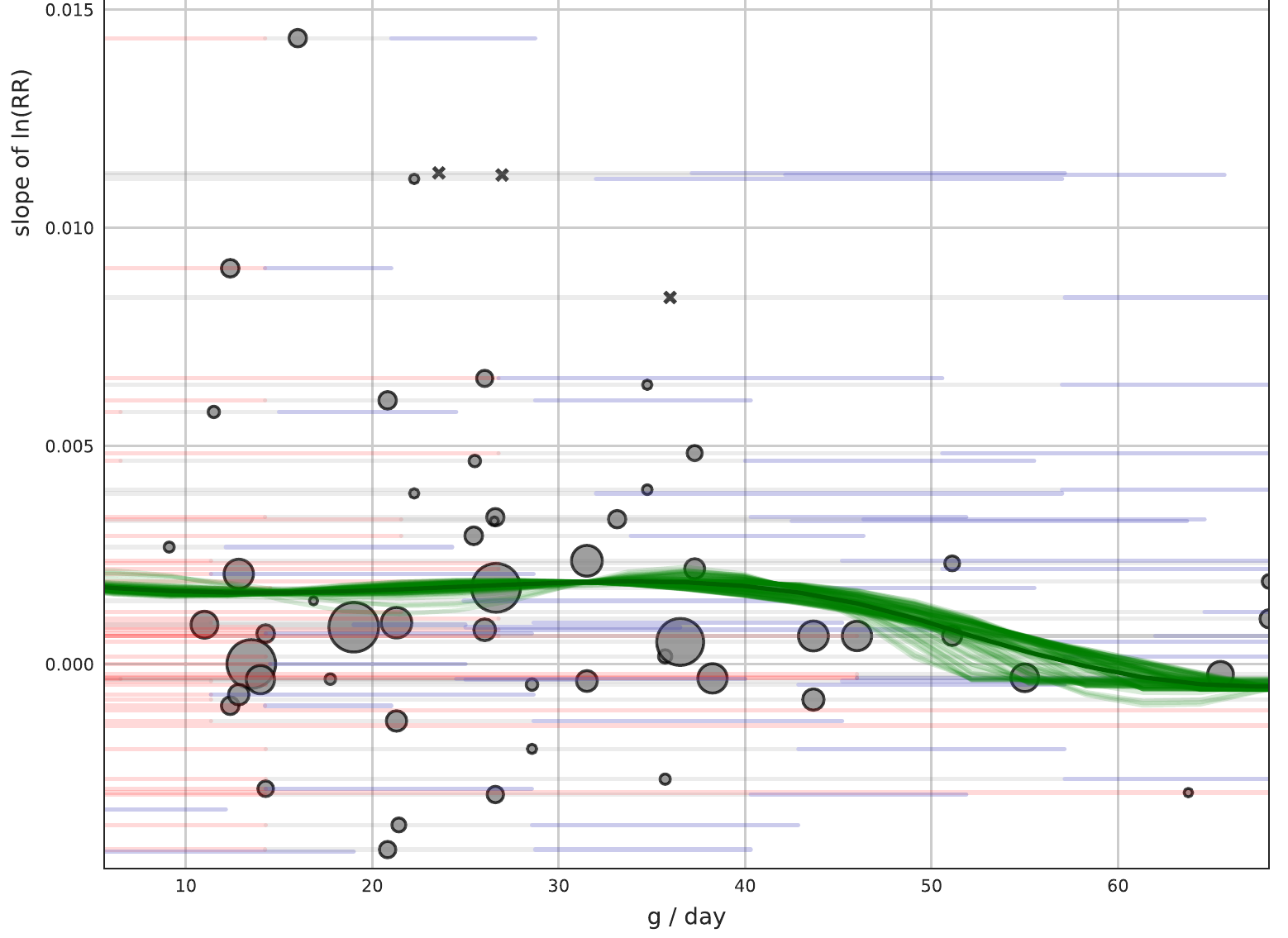}
\caption{Spline fit obtained from observation model~\eqref{eq:indirect_full}. The left panel shows  the inferred dose response mechanism from indirect observations, across multiple spline knot placement options. 
Uncertainty from the spline coefficient fits is plotted in dark gray, with additional uncertainty coming from the random effects shown in light gray. 
 Observations are shown in the right panel. The midpoint between the 
reference and alternative intervals is displayed as a point, with the reference and alternative exposures plotted in blue and red.
\label{fig:meat1}}
\end{figure}
\begin{figure}[h!]
\centering
\includegraphics[width=0.49\textwidth]{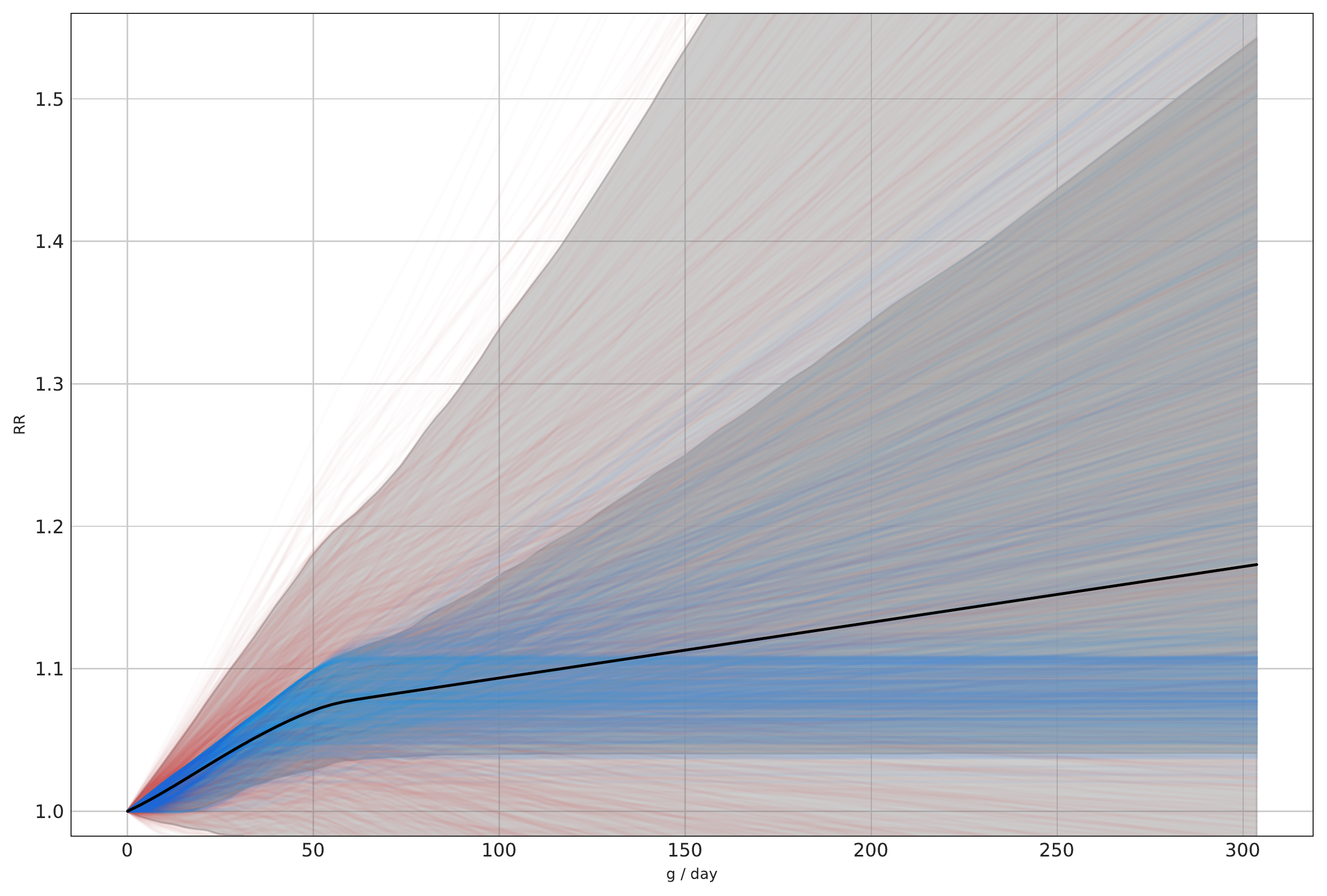}
\includegraphics[width=0.49\textwidth]{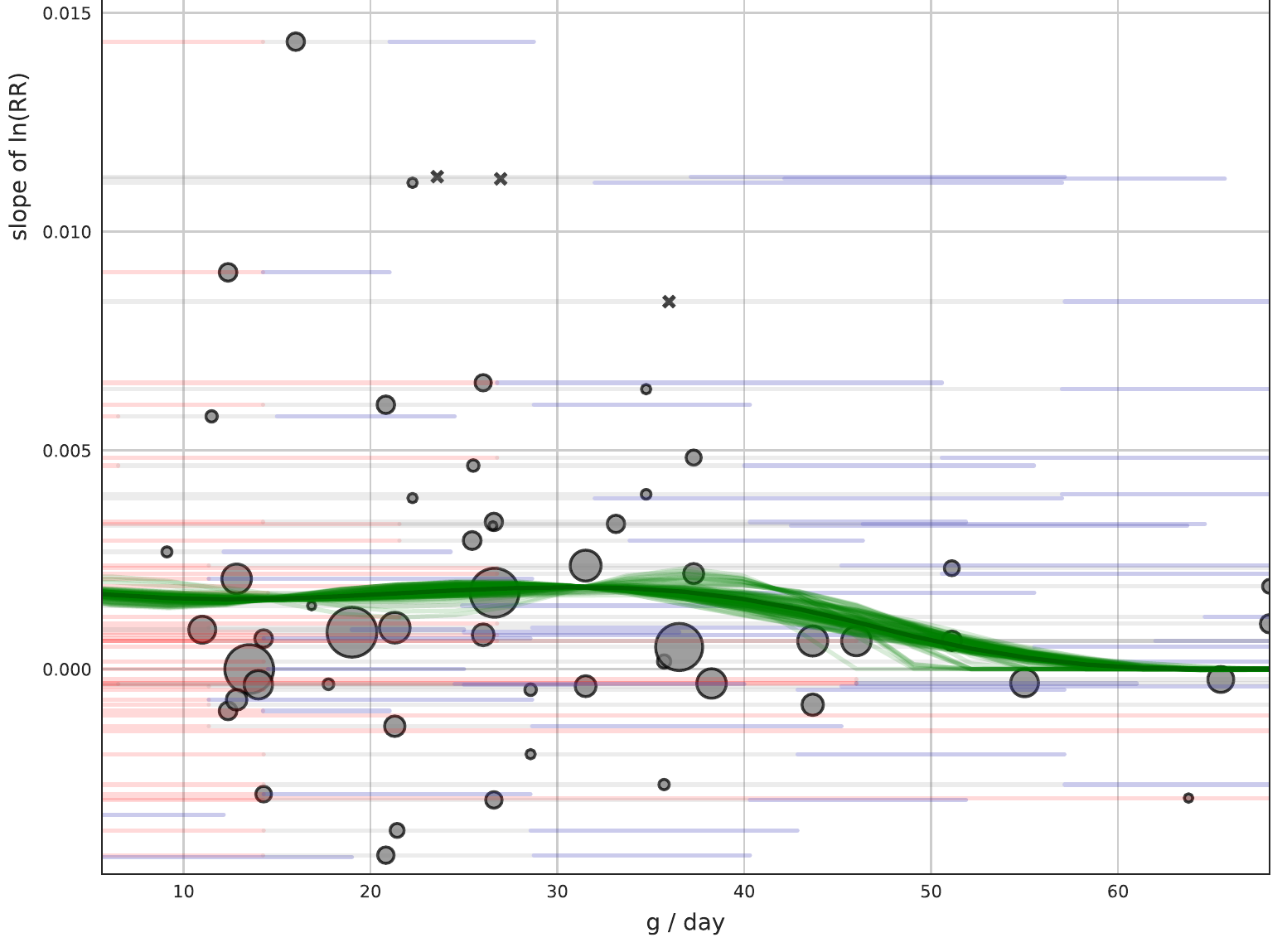}
\caption{Adding monotonicity constraints to example in Figure~\ref{fig:meat1}.  
\label{fig:meat2}}
\end{figure}
As in the smoking example, we model the random effects on the average slopes of the log-relative risks,  use a spline to represent the risk curve, and constrain the risk at zero exposure to be $1$.  
\blue{In contrast to the smoking case, here we must account for both reference and alternative group definitions, so the observation model is given by}
\begin{equation}
    \label{eq:indirect_full}
    \begin{aligned}
    y_{ij} &= 
    \left(\frac{\ln(\langle \bm x_{ij}^1, \bm \beta \rangle) - \ln(\langle \bm x_{ij}^0, \bm \beta\rangle)}{\mbox{alt exposure}_{ij} - \mbox{ref exposure}_{ij}} + u_i\right) 
    (\mbox{alt exposure}_{ij} - \mbox{ref exposure}_{ij}) + \epsilon_{ij} \\
    & = \ln(\langle \bm x_{ij}^1, \bm \beta \rangle) - \ln(\langle \bm x_{ij}^0, \bm \beta\rangle) +  (\mbox{alt exposure}_{ij} - \mbox{ref exposure}_{ij})u_i + \epsilon_{ij}.
    \end{aligned}
\end{equation}
with \blue{$y_{ij}$ the log relative risk,} $\bm x_{ij}^1$ and $\bm x_{ij}^0$ computed using spline basis matrix for alternative exposure$_{ij}$
and reference exposure$_i$,
(see Section~\ref{sec:spline_basics}), $u_i$ the random effect for study $i$, and 
$\epsilon_{ij} \sim N(\bm 0, \sigma_{ij}^2)$ the variance reported by the $i$th study for its $j$th point.

 \LimeTr  is the only package that can infer the nonlinear dose-response relationship in this example using heterogeneous observations of ratios of average risks across variable exposures. The meta-analysis to obtain the risk curve in Figure~\ref{fig:meat1} integrates results from 14 prospective cohort studies\footnote{Variation in risk here is much lower than for smoking, so we show relative risk instead of log relative risk.}.
 An ensemble of spline curves was used rather than a single choice of knot placement. 
 The cone of uncertainty coming from the spline coefficients is shown in dark gray in the left panel, with additional uncertainty from random effects heterogeneity shown in light gray.  
 
 The right panel of Figure~\ref{fig:meat1} shows the data fit. Each point represents a ratio between risks integrated across two intervals, so in effect on four points 
 that define these intervals. We choose to plot the point at the midpoint defined by these intervals for the visualization. As in other plots, the radius of each point is inversely proportional to the standard deviation reported for it by the study. The derivatives of each spline curve fit are plotted in green so that they can be compared to these average ratios on log-derivative space.

After studying the relationship and the data in Figure~\ref{fig:meat1}, 
 it is hard to justify forcing a monotonic increase in risk.  However, to test the functionality of \LimeTr , we add this  constraint and show the result in Figure~\ref{fig:meat2}. 
 Our conclusions about the relationship and its strength would change in this example 
if we impose such shape constraints, unlike the example that compares smoking and lung cancer.

\section{Conclusion}
\label{sec:conc}

We have developed a new methodology for robust mixed effects models, and implemented it using the \LimeTr  package.
The package extends the trimming concept widely used 
in other robust statistical models to mixed effects. It solves the resulting problem using a new method that combines a standalone optimizer \texttt{IPopt} with a customized solver
for the value 
function over the trimming parameters introduced in the reformulation. Synthetic examples show that \LimeTr  
is significantly more robust to outliers than available packages for meta-analysis,
and also improves on the performance of packages for robust mixed effects regression/longitudinal analysis.

In addition to its robust functionality, 
\LimeTr  includes additional features that 
are not available in other packages, including 
arbitrary nonlinear functions of fixed effects $\bm f_i(\bm\beta)$, as well as linear and nonlinear constraints. 
In the section that uses empirical data, we have shown how these features can be used to do standard meta-analysis 
as well as to infer nonlinear dose-response relationships from direct and indirect observations 
of complex nonlinear relationships.

\bigskip
\begin{center}
{\large\bf SUPPLEMENTAL MATERIALS}
\end{center}

\begin{description}
      \item[\LimeTr package:] Python package \LimeTr containing code to perform robust estimation of mixed effects models, 
      for both meta-analysis and simple longitudinal analysis. Available online through github: 
      \verb{https://github.com/zhengp0/limetr{      
      \item[Experiments:] Set of script files available online to produce simulated data and run \LimeTr and third party code:
      \verb{https://github.com/zhengp0/limetr/tree/paper/experiments{
      \begin{itemize}
      \item \verb{Settings.R{: R code to Specifies folder structure and simulation parameters. 
      \item \verb{functions.R{: R code for auxiliary functions for simulating data and aggregating results. 
      \item \verb{0_create_sim_data.R{: R code to create simulated data. 
      \item \verb{1_limetr.py{: Python code to run \LimeTr. 
      \item \verb{2_metafor.R{: R code to run \verb{metafor{ package 
      \item \verb{3_robumeta.R{: R code to run \verb{robumeta{ package
      \item \verb{4_metaplus.R{: R code to run \verb{metaplus{ package
      \item \verb{5_lme4.R{: R code to run \verb{lme4{ package
       \item \verb{6_robustlmm.R{: R code to run \verb{robustlmm{ package
      \item \verb{7_heavy.R{: R code to run \verb{heavy{ package
      \end{itemize} 
 \end{description}





\bibliographystyle{plainnat}
\bibliography{rss}

\end{document}